%% file: main.tex
\documentclass[num-refs]{wiley-article}
\usepackage{booktabs}
\usepackage{siunitx}
\DeclareSIUnit{\instruction}{instruction}
\DeclareSIUnit{\float}{float}
\DeclareSIUnit{\characters}{characters}
\DeclareSIUnit{\Mfps}{\mega\float\per\second}
\usepackage{microtype}
\usepackage{lipsum}
\PassOptionsToPackage{hyphens}{url}
\usepackage{hyperref}
\usepackage{multirow}
\usepackage{xcolor}
\hypersetup{
    colorlinks,
    linkcolor={red!50!black},
    citecolor={blue!50!black},
    urlcolor={black}
}
\newcommand*{\restartrowcolors}{%
  \ifhmode\unskip\fi
  \vadjust{%
    \global\rownum=0 %
  }%
}
\usepackage{tikz}
\usetikzlibrary{shapes,arrows,arrows.meta,positioning}

\usepackage{pgfplots}
\pgfplotsset{compat=1.17}
\usepackage{subcaption}

\usepackage{textcomp}

\usepackage{listings}
\usepackage{pythonhighlight}

\usepackage[utf8]{inputenc}

\title{Converting Binary Floating-Point Numbers to Shortest Decimal Strings: An Experimental Review}

\author[1]{Ja\"el Champagne Gareau}
\author[1]{Daniel Lemire}

\corraddress{Daniel Lemire, Universit\'e du Qu\'ebec (TELUQ), Montreal, Quebec, H2S 3L5, Canada}
\corremail{daniel.lemire@teluq.ca}

\fundinginfo{Natural Sciences and Engineering Research Council of Canada, Grant Number: RGPIN-2024-03787\\Fonds de recherche du Québec, \url{https://doi.org/10.69777/361128}}

\affil[1]{Universit\'e du Qu\'ebec (TELUQ), Montreal, Quebec, H2S 3L5, Canada}
\date{}

\begin{document}
\maketitle

\begin{abstract}
When sharing or logging numerical data, we must convert binary floating-point
numbers into their decimal string representations. For example, the number $\pi$
might become \texttt{3.1415927}. Engineers have perfected many algorithms for
producing such accurate, short strings. We present an empirical comparison
across diverse hardware architectures and datasets. Cutting-edge techniques like
Schubfach and Dragonbox achieve up to a tenfold speedup over Steele and White's
Dragon4, executing as few as 210 instructions per conversion compared to
Dragon4's 1500--5000 instructions. Often per their specification, none of the
implementations we surveyed consistently produced the shortest possible
strings---some generate outputs up to 30\% longer than optimal. We find that
standard library implementations in languages such as C++ and Swift execute
significantly more instructions than the fastest methods, with performance gaps
varying across CPU architectures and compilers.
We suggest some optimization targets for future research.
\end{abstract}

\section{Introduction}

Processor vendors have adopted 32-bit and 64-bit IEEE~754 floating-point
numbers. Consequently, we typically represent numbers as IEEE~754
floating-point numbers in software. The corresponding types in Java, C, C\# and
C++ are \texttt{float} and \texttt{double}. JavaScript uses the 64-bit IEEE~754
floating-point format as its default number data type. These numbers take the
form of a fixed-precision integer (the significand\footnote{The term
\emph{mantissa} is sometimes used as a synonym for \emph{significand}, but this
usage is discouraged by the IEEE~754 standard, which reserves the term
\emph{significand} to denote the fractional component of a floating-point
number. In contrast, \emph{mantissa} historically referred to the fractional
part of a logarithm.}) multiplied by a power of two ($m \times 2^p$). For
example, a 32-bit floating-point value approximating the constant $\pi$ is
$13176795 \times 2^{-22}$. We often convert these binary values into decimal strings:
\begin{itemize}
  \item when we serialize data to text formats like CSV, JSON or YAML;
  \item when we produce human-readable logs and telemetry;
  \item when we print numbers in graphical interfaces, spreadsheets and dashboards.
\end{itemize}
Producing the shortest possible string that exactly reproduces the original
value can require hundreds or even thousands of CPU instructions.
Since many applications convert millions or even billions of values in bulk,
that per-value overhead may quickly add up to a substantial performance
bottleneck.

Converting binary floating-point values into decimal strings is largely a matter
of established software practices, yet it remains under-explored in the research
literature. In Section~\ref{sec:problem}, we formally define the shortest-string
conversion problem. In Section~\ref{sec:related}, we survey the principal
algorithmic families. Finally, we present an experimental comparison of key
implementations in Section~\ref{sec:experiments}---ranging from Steele and
White’s 1990 Dragon4 to modern methods like Schubfach and Dragonbox---and
present some directions for future work in Section~\ref{sec:conclusion}.

Taken together, our study offers both a broad empirical perspective and new
methodological insights. Our main contributions are the following:
\begin{itemize}
  \item \emph{A systematic empirical evaluation of major conversion algorithms},
    spanning both dominant CPU families (x86-64 and ARM/AArch64) and an
    expanded, openly available benchmark suite including mostly real-world datasets.
  \item \emph{New measurements of output behavior}, including the first
    detailed characterization of end-to-end string lengths---which often differ
    from minimal significand lengths---and instruction-level metrics that
    isolate algorithmic cost from microarchitectural throughput.
  \item \emph{A reassessment of existing benchmarking practices}, identifying
    methodological limitations in prior evaluations.
\end{itemize}

\section{Problem Definition}%
\label{sec:problem}

Although IEEE~754 floating-point types are the default choice for representing
real numbers in software---and are widely supported by commodity processors
and programming languages---their use can lead to non-obvious difficulties. In
this section, we formalize the problem and outline its practical implications.
Table~\ref{tab:commmonieee} describes the bit layouts of 32- and 64-bit
floating-point numbers. The IEEE~754 formats dedicate a bit for the sign:
accordingly, we can distinguish between $-0$ and $0$. A positive \emph{normal}
double-precision floating-point number is a binary floating-point value whose
significand is represented with 53 bits of precision: 52 bits are explicitly
stored, while the leading 1~bit is implicit (not physically stored in memory).
For example, the value with a significand of $1.011_2$ would be stored as an
implicit leading $1$ and explicit bits $011$. As such, the significand can be
seen as a 53-bit integer~$m$ in the interval $[2^{52}, 2^{53})$ but interpreted
as a number in $[1, 2)$ by dividing it by $2^{52}$. The 11-bit exponent $p$
ranges from \num{-1022} to \num{+1023}~\cite{10.1145/103162.103163}. Values
smaller than $2^{-1022}$ are called \emph{subnormal} values: their special
exponent code has the value $2^{-1022}$ and the significand is then interpreted
as a value in $[0, 1)$. We can uniquely identify a 64-bit number using a
17-decimal-digit representation, although fewer digits are often needed. The
32-bit numbers are similarly defined, with a 24-bit significand $m$ and an 8-bit
exponent $p$ ranging from \num{-126} to \num{+127}. Numbers smaller than
$2^{-126}$ are represented using a subnormal format. We have that 9~digits are
sufficient to uniquely identify a 32-bit number.

\begin{table}
  \centering
  \caption{Common IEEE~754 binary floating-point numbers}
  \label{tab:commmonieee}
  \begin{tabular}{cccc}
    \toprule
    name     & exponent bits & significand (stored) & decimal digits (exact) \\\midrule
    binary64 & 11~bits       & 53~bits (52~bits)    & 17 \\
    binary32 & 8~bits        & 24~bits (23~bits)    & 9 \\
    \bottomrule
  \end{tabular}\restartrowcolors
\end{table}

\begin{figure}
  \centering
  \begin{tikzpicture}[
      box/.style={rectangle, draw, rounded corners, minimum height=3em, minimum width=2.1cm, align=center, fill=blue!10},
      convbox/.style={rectangle, draw, rounded corners, minimum height=3em, minimum width=5.6cm, align=center, fill=orange!10},
      arrow/.style={-Stealth, thick},
      font=\footnotesize
    ]

    \node[box, minimum width=2.1cm] (sign) at (0,0) {Sign bit\\1 bit\\0 or 1\\(-0 or +0)};
    \node[box, minimum width=3.5cm, right=0.35cm of sign] (exponent) {Exponent\\11 bits\\$p \in [-1022, +1023]$};
    \node[box, minimum width=5.6cm, right=0.35cm of exponent] (significand) {Significand\\53 bits (1 implicit + 52 explicit)\\$m \in [2^{52}, 2^{53})$, interpreted in $[1, 2)$\\Ex: $1.011_2$ stored as 011};

    \node[convbox, minimum width=2.1cm, below=1cm of sign] (plusminus) {+/-};
    \node[convbox, minimum width=5.6cm] (conv) at (5.95,-3) {Convert to decimal}; 
    \node[convbox, minimum width=8.4cm, below=2cm of conv] (string) {Decimal string representation};

    \draw[arrow] (sign.south) -- (plusminus.north);
    \draw[arrow] (exponent.south) -- ++(0,-0.5) -| (conv.north);
    \draw[arrow] (significand.south) -- ++(0,-0.5) -| (conv.north);
    \draw[arrow] (plusminus.south) to[out=-90,in=180] ([xshift=-2cm]string.north); 
    \draw[arrow] (conv.south) -- (string.north);
  \end{tikzpicture}
  \caption{Conversion of a 64-bit number to a string}%
  \label{fig:tostring}
\end{figure}
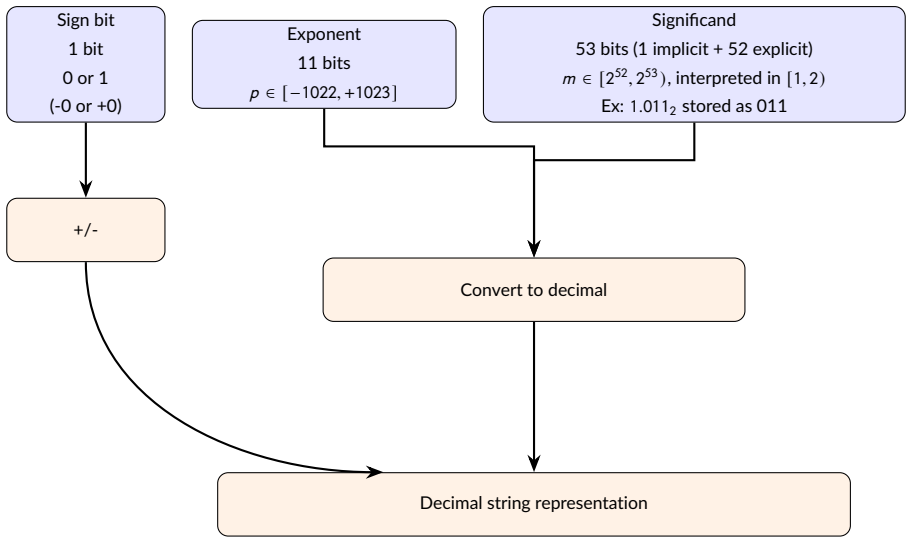

String representations are typically in decimal format. Converting a binary
floating-point number to a decimal string is usually done in three steps:
\begin{enumerate}
  \item Extracting the sign bit, exponent, and significand;
  \item Converting the binary significand and exponent to their decimal counterparts;
  \item Generating the string representation of the resulting decimal number.
\end{enumerate}
See Fig.~\ref{fig:tostring}.
The first step---extracting the three fields from the bit pattern---is straightforward and leaves little room for algorithmic
innovation. Consequently, most research has focused on the second step: the
conversion of a value from a base-2 representation ($m \times 2^p$) to a base-10
representation ($w \times 10^q$). This conversion involves determining the decimal
significand $w$ and exponent $q$ from the binary significand $m$ and exponent
$p$ by solving the equation $m \times 2^p = w \times 10^q$~\cite{10.1145/362851.362887}.

An exact solution is always possible in this direction (from binary floating-point to decimal string). For $p \geq 0$, setting
$q = 0$ and $w = m \times 2^p$ suffices, since any positive power of two is
exactly representable in base ten. For $p < 0$, we can set $q = p$ and $w = m
\times 5^{-q}$, making use of the identity $2^p = 10^p / 5^p$.

However, the converse is not generally true: most decimal fractions cannot be represented
exactly as binary floating-point numbers. For example, $0.1$ cannot be expressed
as $m \times 2^p$ with integer $m$, since $m = 2^{-p-1}/5$ would require a power
of two to be divisible by $5$, which is impossible.

After accounting for special cases ($\pm0$, $\pm\infty$, NaN, and subnormals)
and ignoring the sign bit, the central mathematical problem becomes finding the
smallest integer $w$ such that $w \times 10^q$ maps to the original binary
floating-point value and to no other representable value. In other words, the
decimal representation should unambiguously represent the intended
floating-point value rather than to any adjacent representable value.
By finding the smallest integer $w$, we seek to minimize the number of
\emph{significant digits} in the decimal string. Significant digits are the
digits in a number that contribute to its precision, including all non-zero
digits and zeros between non-zero digits. For example, the number 1100 is
generally considered to have two significant digits, namely the two 1s, as the
trailing zeros are ambiguous and not counted without explicit clarification.
We often use scientific notation, where a number is expressed as a coefficient
multiplied by a power of 10, such as $ a \times 10^b $ or in ``\texttt{E}''
notation as \texttt{aEb}, to clearly define the significant digits. For
instance, writing 1100 as \( 1.1 \times 10^3 \) or {1.1E3} indicates two
significant digits, ensuring precision is unambiguous.

To illustrate these ideas, consider the following example using the number $\pi$. Its
closest 32-bit floating-point approximation (in binary) is $13176795 \times 2^{-22}$.
The corresponding exact decimal representation is: $$31415927410125732421875 \times 10^{-22}.$$
However, it is not necessary to print the entire integer part in most
applications. For 32-bit floating-point numbers, nine decimal digits are always
sufficient to uniquely distinguish each value. Therefore, we can round this
exact decimal to nine digits, giving $314159274 \times 10^{-8}$ for our
candidate decimal representation. In this case, we truncate down to nine
digits: since the next digit is `1', we do not round up. However, it is actually
possible to use a shorter decimal: $31415927 \times 10^{-7}$. This shorter form
is still closer to $13176795 \times 2^{-22}$ than to any other 32-bit float, and
is therefore the shortest unique decimal representation.

After determining the decimal significand and exponent, the final step is to
generate the corresponding string representation (e.g., producing
\texttt{3.1415927} from $w$ and $q$ in $w \times 10^q$). This process is not
always cleanly separable from the earlier computation: some algorithms combine
string generation and significand calculation, while others treat them as
distinct stages. Typically, the decimal digits of an integer value are obtained
by repeatedly extracting the least significant digit ($w \bmod 10$) and dividing
by ten, with various optimizations possible. For example, a common optimization
is to proceed by pairs of digits ($w \bmod 100$) and use a table to map the
value in $[0,100)$ to a pair of characters (e.g., the value 10 maps to the
string `\texttt{10}'). However, even if the smallest significand has been found
in the previous step, further care is needed if we want to ensure that the
output string is as short as possible. For instance, both \texttt{9.9E1} and
\texttt{99} represent the same value with two significant digits, but the latter
is shorter and arguably preferred; scientific notation should only be used when
it produces a shorter overall string.

It is crucial to distinguish between two closely related but conceptually different
objectives in float-to-string conversion:
\begin{itemize}
  \item (1) \emph{Minimal decimal significand.} Most algorithms
    aim to compute the shortest decimal significand $w$ such that $w \times
    10^q$ round-trips to the original binary floating-point value. This solves
    step~2 of the conversion pipeline.
  \item (2) \emph{Minimal printed string.} We might instead wish to find the
    shortest \emph{character string} that round-trips, including the choice
    between fixed-point and scientific notation and the placement of the decimal point.
\end{itemize}
These two goals coincide for many values but are not equivalent. Two
representations with the same number of significant digits may differ in total
length depending on formatting choices. For example, the binary64 value
$12000000000$ has a shortest significand ``12'', but the shortest overall
string is ``12e9'' (4 characters), whereas the canonical scientific form
``1.2e10'' contains six characters. Consequently, existing algorithms typically
optimize~(1)---the computation of the minimal significand---but not necessarily~(2).

Representation subtleties can significantly affect float-to-string conversions.
Consider the integer \num{2150000128}.
We can represent it exactly as a 32-bit floating-point number ($8398438\times 2^{8}$). The previous 32-bit floating-point number is \num{2149999872} or $8398437\times 2^{8}$. The number \num{2150000000} falls between these two 32-bit floating-point values.
 When a decimal value lies exactly halfway between two floats,
as here, IEEE~754 requires rounding to the value whose least significant bit is
zero (round-to-even).\footnote{Round-to-even, also known as banker's rounding,
is a method for rounding decimal numbers to reduce bias in calculations, often
used in financial or statistical contexts. When a number is exactly halfway
between two integers (e.g., 2.5 or 3.5), instead of always rounding up (as in
standard rounding), round-to-even rounds to the nearest even integer. This
method evens-out the frequency of rounding up and down over many operations.}
Hence, the string \texttt{2150000000} is parsed as the 32-bit float value \texttt{2150000128}. Thus, when converting the 32-bit floating-point number
\texttt{2150000128} to a string, we should write \texttt{2.15e9} to minimize
the number of digits in the significand and the string length. Such cases underscore the need for
careful handling of rounding and highlight edge cases that conversion algorithms
must address.


\section{Related Work}%
\label{sec:related}

This section reviews the main algorithmic families for shortest float‐to‐string
conversion, then pinpoints why existing benchmarks leave important questions
unanswered.

\subsection{Existing Algorithms}

The conversion of binary floating-point numbers to their decimal string
representation has been a subject of research over the years.
However, there are relatively few scholarly contributions.


A crucial but long-overlooked contribution was made by Coonen in his 1980
technical report~\cite{coonen1980} and extended in Chapter 7 of his 1984 PhD
Thesis~\cite{coonen1984}. Coonen was the first to articulate the modern view of
the correct-rounding, formulating the requirement that a printed decimal must
round back to the original binary floating-point value and introducing the
interval-based reasoning later reused in algorithms such as Grisu and Ryū.
Although this work circulated primarily as a technical report and was not widely
cited in early literature, its importance has since been recognized---most
prominently by Loitsch, who identifies Coonen's work as the earliest description
of a correct floating-point–to–decimal algorithm.


In 1990, Steele and White~\cite{steele1990print} authored the foundational
article that brought these ideas into the programming-language community and
established the Dragon family of algorithms. They report that their work was
originally conducted in the late 1970s and circulated informally for many
years before publication. Their Dragon4 algorithm can thus be viewed as a
practical successor to the conceptual framework introduced by Coonen. The
publication of the IEEE~754 standard in 1985~\cite{ieee1985floating} occurred
after many implementations were already in use, and Steele and White’s paper
was the first widely disseminated description of a fully practical
correct-rounding and shortest-decimal algorithm.

In their paper, Steele and White introduce core objectives for the
conversion of floating-point values to decimal representation. They distinguish
between fixed-format outputs---where the number of digits is predetermined
(e.g., always printing five digits)---and free-format outputs, where the
algorithm determines the number of digits dynamically.

For the free-format case, they propose three essential properties that a correct
conversion algorithm from binary floating-point values to decimal should satisfy:
\begin{enumerate}
  \item \emph{No loss of information}: Converting back the generated decimal
    representation must recover the original floating-point value (i.e.,
    round-trip conversion is the identity function).
  \item \emph{No extra information}: The generated decimal representation must
    not contain extraneous digits (i.e., we want the minimal significand possible).
  \item \emph{Correct rounding}: Multiple values may satisfy the first two
    properties. Among these, the algorithm should produce the one closest to
    the original floating-point value, using round-to-even in the case of a tie.
\end{enumerate}


To address these objectives, Steele and White first present a practical, though
approximate, algorithm for generating free-format output: Dragon2 (see
Fig.~\ref{fig:dragon2} for a Python sketch). The algorithm first rescales the
floating-point value so that it falls within the interval $[0,1)$, and then uses
floating-point arithmetic to compute the digits. As the authors remark, the
algorithm is not accurate due to its reliance on floating-point arithmetic.
Thus, some values may not be recovered exactly from the output string. Next,
they present an algorithm which they named Dragon4: they omit Dragon3.\footnote{In a retrospective article~\cite{10.1145/989393.989431}, Steele and White explain that the reference to a Dragon has to do with \emph{dragon curves}, a mathematical curiosity. This curve is constructed by combining two types of steps: \emph{Folds} and \emph{Peaks}, whose initials (``FP'') allude to floating-point numbers.}
Several other algorithms in this domain are named after
dragons. Dragon4 is based entirely on integer arithmetic, and it can therefore be
exact (see Fig.~\ref{fig:dragon4} for a sketch in Python). Instead of scaling
the floating-point integer, it scales two integers $R$ and $S$ representing the
floating-point value. The fraction $R/S$ represents the value which is
iteratively scaled to be in a safe subinterval of $[0,1)$. The downside of
Dragon4 is that it may require big-integer arithmetic and several division
operations.

\begin{figure}[htb]
  \centering
  \begin{tabular}{l}
    \inputpythonfile{review/steelewhite90/dragon2.py}
  \end{tabular}\restartrowcolors
  \caption{Python sketch of the Dragon2~\cite{steele1990print} algorithm for non-negative numbers. The algorithm supports conversions between arbitrary bases, not just from base-2 to base-10. Parameters \texttt{b} and \texttt{B} denote the input and output radix, respectively. The parameter \texttt{n} indicates the input precision, in number of significand binary digits (e.g., the default \texttt{n = 24} corresponds to binary32 precision; see Table~\ref{tab:commmonieee}).}%
  \label{fig:dragon2}
\end{figure}

The algorithm produces the digits one by one, stopping when the following digits
would be all zeros. For producing fixed-format outputs, we can generate the
number of digits and then round the final digit if needed. If fewer digits are
produced by the free-format approach than we desire, we pad with zeros. The
rounding up may then require changing already produced digits (e.g., if the
digit 9 is rounded up). Steele and White observe that this can be avoided by an
algorithm that already produces the correct digits, knowing how many are needed.

A significant amount of computation in Dragon4's algorithm is spent scaling the
values: the computation of the constant $k$ in Fig.~\ref{fig:dragon4} requires
many operations. For example, a floating-point value such as \texttt{1e-300}
requires over 299~iterations, each of which entails three multiplications.

\begin{figure}[htb]
  \centering
  \begin{tabular}{l}
    \inputpythonfile{review/steelewhite90/dragon4.py}
  \end{tabular}\restartrowcolors
  \caption{Python sketch of the Dragon4~\cite{steele1990print} algorithm for non-negative numbers.}%
  \label{fig:dragon4}
\end{figure}


Following Steele and White, Gay~\cite{gay1990correctly} presented methods for
accurately converting between binary floating-point numbers and decimal strings,
with a specific focus on ensuring correct rounding. He describes the
\texttt{dtoa} function, which implements various modes such as free-format
(shortest-string) and fixed-format output. A key insight is that instead of
using the relatively slow Steele-White algorithm to compute the parameter $k$
(see Fig.~\ref{fig:dragon4}), we can use a faster floating-point approach,
correcting it if needed. Further, Gay observes that when the number of
significant digits is sufficiently small, we may use a faster algorithm based on
floating-point arithmetic, producing an exact result with fewer computations. We
may also check whether the floating-point value is an integer, in which case a
conversion to an integer value and its printing might be faster. Gay's
\texttt{dtoa} function is several times faster than Dragon4. Fig.~\ref{fig:gay}
illustrates \texttt{dtoa} using Python code. For simplicity, it omits the path
where the \texttt{dtoa} function uses floating-point arithmetic. Burger and
Dybvig~\cite{10.1145/249069.231397} contributed techniques similar to Gay's by
developing a scale estimator.

\begin{figure}[htb]
  \centering
  \begin{tabular}{l}
    \inputpythonfile{review/gay90/dtoa.py}
  \end{tabular}\restartrowcolors
  \caption{Python sketch of the Gay's \texttt{dtoa} function for non-negative numbers.}%
  \label{fig:gay}
\end{figure}




Loitsch~\cite{10.1145/1806596.1806623} explored an integer-based approach to
achieve quick and accurate conversions. Their approach uses a combination of
precomputed powers of ten and a \texttt{diy\_fp} (do-it-yourself floating-point)
representation (a 64-bit integer coupled with an integer for the exponent) to
approximate the decimal output quickly. The precomputed powers of ten must hold
roughly 635~precomputed values for 64-bit floating-point numbers. Starting from
a suboptimal Grisu algorithm, they have developed an algorithm called Grisu2
which is often, but not always, able to produce the shortest representation.
They have also developed another algorithm called Grisu3 which detects when its
output may not be the shortest. Loitsch proposes falling back on another
printing algorithm like Dragon4 when Grisu3 fails.


Andrysco, Jhala, and Lerner~\cite{10.1145/2837614.2837654} introduced the Errol
algorithm for printing floating-point numbers, aiming for both correctness and
speed. Initially, they claimed it was significantly faster than Grisu3; however,
they later acknowledged in their repository that this evaluation was flawed:
\emph{``Our original evaluation of Errol against the prior work of Grisu3 was
erroneous\dots~Corrected performance measurements show a 2x speed loss to
Grisu3.''} We tested their
implementation,\footnote{\url{https://github.com/marcandrysco/Errol}} but found
it contained unsafe code, so we do not consider Errol further.


Adams~\cite{adams2018ryu,adams2019ryu} introduced the Ryū algorithm, which, like
Grisu3, targets fast and correct printing of floating-point numbers, but with
the crucial advantage of always guaranteeing correct output, thus eliminating
the need for a fallback algorithm. The correctness of Ryū relies on several key
elements described in the 2018 paper~\cite{adams2018ryu}. First, Ryū decodes a
floating-point number into a unified representation that handles both normalized
and subnormal cases, written as $f = (-1)^s \cdot m_f \cdot 2^{e_f}$, where
$m_f$ is an unsigned integer. The algorithm then calculates the interval of
decimal values that would be decoded back to the original binary float, by
determining the midpoints to the immediately smaller and larger representable
floating-point numbers. These midpoints define an interval $[u, w)$, scaled by
$2^{e_2}$, within which any value will round to the original float. This
interval is then converted into decimal base, yielding an interval $[a, c) \cdot
10^{e_{10}}$. The core digit-generation step identifies the shortest decimal
representation within this interval by iteratively removing digits from right to
left. At each step, it ensures that the shortened decimal string still lies
within the valid interval, thus guaranteeing that parsing this string returns
the original float. To handle rounding modes correctly, Ryū analyzes the prime
factorization of the significand to decide whether the result should be rounded
up or down. This process is carried out using only fixed-precision integer
operations, which avoids costly high-precision multiplications. A major
contribution of Ryū is a reduction in the number of bits required for
intermediate computations. Instead of converting the entire binary value to
decimal in one high-precision operation, Ryū combines incremental decimal
conversion with its digit-generation loop. This allows the use of smaller
intermediate values, aided by precomputed lookup tables for multipliers such as
$\lfloor 2^k / 5^q \rfloor + 1$ or $\lfloor 5^{-e_2-q} / 2^k \rfloor$, which
enable efficient and accurate calculation. In a follow-up
paper~\cite{adams2019ryu}, Adams extends Ryū to create Ryū Printf, supporting
the \texttt{\%f}, \texttt{\%e}, and \texttt{\%g} formats with
runtime-configurable precision. To maintain efficiency, Ryū Printf introduces a
segmentation approach, converting the significand into segments of decimal
digits (e.g., 9 digits per segment for 32-bit integers). Each segment is
computed independently, so computational cost is linear in the number of digits
generated, rather than superlinear as in naive approaches.


\begin{table}[htpb]
  \centering
  \caption{Overview of binary floating-point to string algorithms. A technique is considered exact if the obtained string representation is always sufficient to recover the original binary floating-point number.}%
  \label{tab:characteristics}
  \begin{tabular}{lccc}\toprule
    Technique           & Source                                  & Publication Date & Exact \\ \midrule
    Dragon2             & Steele and White~\cite{steele1990print} & 1990             & No \\
    Dragon4             & Steele and White~\cite{steele1990print} & 1990             & Yes \\
    \texttt{dtoa}       & Gay~\cite{gay1990correctly}             & 1990             & Yes \\
    Grisu               & Loitsch~\cite{10.1145/1806596.1806623}  & 2010             & No \\
    Grisu2              & Loitsch~\cite{10.1145/1806596.1806623}  & 2010             & No \\
    Grisu3 + Dragon4    & Loitsch~\cite{10.1145/1806596.1806623}  & 2010             & Yes \\
    Ryū                 & Adams~\cite{adams2018ryu,adams2019ryu}  & 2018             & Yes \\
    Schubfach           & Giulietti~\cite{schubfach}              & 2020 (informal)  & Yes \\
    Grisu-Exact         & Jeon~\cite{grisu-exact}                 & 2020 (informal)  & Yes \\
    Dragonbox           & Jeon~\cite{dragonbox}                   & 2022 (informal)  & Yes \\
    \bottomrule
  \end{tabular}\restartrowcolors
\end{table}

Other algorithms not formally published in the peer-reviewed literature include
Giulietti’s Schubfach~\cite{schubfach}, which decomposes a floating-point value
into its significand and exponent, computes tight decimal bounds, and dispatches
into specialized computation paths based on the number’s properties; Jeon’s
Dragonbox~\cite{dragonbox}, inspired by the Ryū algorithm; and Jeon’s earlier
Grisu-Exact~\cite{grisu-exact}, a Grisu-family variant that guarantees shortest,
correctly rounded outputs. See Table~\ref{tab:characteristics} for a summary of
each algorithm’s characteristics. For each entry, we list the original source
and indicate whether it produces exact free-format strings (i.e., the 
strings that permit error-free round-trip parsing). The \emph{Grisu3 + Dragon4}
technique denotes Grisu3 with a Dragon4 fallback.


\subsection{Limitations of Prior Benchmarks}%
\label{sec:limitations-benchmarks}

Several open-source benchmarks have been published to compare the performance of
algorithms converting floating-point numbers to decimal strings. Notable
examples include Yip’s \texttt{dtoa-benchmark}\footnote{\url{https://github.com/miloyip/dtoa-benchmark}},
Lugowski’s \texttt{parse-bench}\footnote{\url{https://github.com/alugowski/parse-bench}},
and Bolz’s \texttt{Drachennest}\footnote{\url{https://github.com/abolz/Drachennest}}. Many
algorithm papers---or their associated repositories---also report benchmark
results. Table~\ref{tab:benchmarks} summarizes the most relevant recent
benchmarks, the algorithms they compare, and the test data they use.

\begin{table}[htpb]
  \centering
  \caption{Overview of existing benchmarks and their limitations}%
  \label{tab:benchmarks}
  \begin{tabular}{p{2.1cm}p{4cm}p{6.5cm}}\toprule
    Benchmark & Compared algorithms & Test data \\ \midrule
    \texttt{dtoa-benchmark} & Mostly Grisu family methods & Generates 1\,000 random 64-bit values; \\
    \texttt{parse-bench} & Principally Ryū, Dragonbox and \texttt{std::to\_chars} & Three hard-coded values (\texttt{123456}, \texttt{1}, \texttt{333.323}) repeated many times; \\
    \texttt{Drachennest} & Grisu3, Ryū, Schubfach, Dragonbox, \texttt{std::to\_chars} & Random doubles in $[1,2]$; random doubles in $[10^k,10^{k+1}]$; random doubles in $[0; 10^{10}]$; random 64-bit bit-patterns; \\
    \texttt{Ryū's paper} & C impl.\ of Ryū v.\ double-conversion (Grisu3 impl.); Java impl.\ of Ryū v.\ OpenJDK's native formatter and Jaffer’s variant~\cite{jaffer2018} & Random sampling (Mersenne Twister) of 1,000 32- and 64-bit values interpreted as floats; \\
    \texttt{Dragonbox's preprint} & Ryū, Grisu-Exact, Schubfach, Dragonbox & 100,000 random numbers (random significand and exponent) measured 1,000 times each; 1,000,000 uniformly generated floats measured 1,000 times each; \\
    \bottomrule
  \end{tabular}\restartrowcolors
\end{table}

A key limitation of these benchmarks is that they do not cover all important
algorithms listed in Table~\ref{tab:characteristics}, nor do they evaluate
standard or third-party libraries from other languages (e.g., Google's
double-conversion or Swift's C++ \texttt{dtoa}). Another common limitation is
the restricted hardware and compiler environments used for evaluation: for
example, the Dragonbox preprint benchmarks were run solely on an Intel i7-7700HQ
using Clang-cl. In addition, none of these benchmarks measure the length of the
generated strings, even though string length is crucial for fair
comparison---algorithms producing shorter outputs may not be directly comparable
to those producing longer ones.

The benchmark datasets themselves also have important shortcomings. All but
Ryū's paper consider only 64-bit floating-point numbers. Nevertheless, 32-bit
floating-point values remain widely used in applications where reduced memory
footprint or bandwidth efficiency matter. Benchmarks that focus exclusively on
64-bit numbers therefore miss important practical scenarios, e.g., in mobile
applications, GPUs, and embedded systems. More importantly, all rely on
synthetic data, whereas uniformly distributed floating-point numbers are rarely
encountered in practice. In real-world applications---such as telemetry,
finance, or science---floating-point values typically exhibit non-uniform
distributions, with some values much more frequent than others. Taken together,
these methodological gaps---incomplete algorithm and library coverage; lack of
real-world datasets and 32-bit numbers; limited hardware and compiler
configurations; and absence of any evaluation of output string
lengths---motivate our empirical study.

\subsection{Parsing Numbers}

A tangentially related problem consists of parsing strings to recover binary
floating-point numbers. The early work was conducted by
Clinger~\cite{10.1145/93548.93557,10.1145/989393.989430}. His work describes an
accurate decimal to binary conversion. Gay~\cite{gay1990correctly} improved upon
Clinger's work by introducing several new optimizations.
Lemire~\cite{lemire2021number} provides a significantly (e.g., $4\times$) faster
approach by observing that, in the common case, the significand fits in a 64-bit
word and only needs to be multiplied by (at most) a 128-bit integer. Mushtak and
Lemire completed the work by showing that no fallback is necessary: the core
algorithm is guaranteed to succeed~\cite{MushtakLemire}.

\section{Experiments}%
\label{sec:experiments}

This section details our experimental setup and methodology, designed to address
the gaps identified previously. We first describe the systems used for
benchmarking, the datasets employed, and the algorithms and libraries tested.
The subsequent subsections present the results and main findings of our experiments.

\subsection{Systems}%
\label{sec:systems}

Our benchmarks were executed on the systems listed in Table~\ref{tab:test-cpus}.
The Apple M4 Max results were obtained on a MacBook Pro (2024), while all other
systems---except the Ryzen~9900X---were hosted on Amazon Web Services (AWS).
For most systems, we compared binaries compiled with both the GNU C++ compiler
(\texttt{g++}) and the LLVM Clang compiler (\texttt{clang++}), using the
corresponding standard libraries (\texttt{libstdc++} and \texttt{libc++}). On
the Apple M4 Max, only Clang was used. Unless otherwise noted, we used
\texttt{g++} version~13 and \texttt{clang++} version~18. On the Apple M4 Max, we
used Apple Clang~17. On the AMD Ryzen~9900X, we used \texttt{g++}-15 and \texttt{clang++}-20.

To our knowledge, this is the first evaluation to include both contemporary
x86-64 (Zen 2--5, Ice Lake, Sapphire Ridge) and ARM/AArch64 (M4 Max, Neoverse
N1/V1/V2) architectures in a unified experimental framework.

 \begin{table}
   \caption{Systems used for benchmarking}%
   \label{tab:test-cpus}
   \centering
   \begin{minipage}{\textwidth}
     \centering
     \begin{tabular}{cccc}\toprule
       Processor         & Frequency                & Microarchitecture             & Memory               \\ 
       \midrule
       Apple M4 Max      & \SIrange{4.4}{4.5}{\GHz} & unnamed (aarch64, 2024)       & LPDDR5X (7500\,MT/s) \\ 
       AMD Ryzen 9 9900X & \SIrange{4.4}{5.6}{\GHz} & Zen 5 (x86-64, 2024)          & DDR5 (6000\,MT/s)    \\ 
       AWS Graviton 2    & \SIrange{2.5}{2.5}{\GHz} & Neoverse N1 (aarch64, 2019)   & DDR4 (3200\,MT/s)    \\ 
       AWS Graviton 3    & \SIrange{2.6}{2.6}{\GHz} & Neoverse V1 (aarch64, 2022)   & DDR5 (4800\,MT/s)    \\ 
       AWS Graviton 4    & \SIrange{2.8}{2.8}{\GHz} & Neoverse V2 (aarch64, 2024)   & DDR5 (5600\,MT/s)    \\ 
       AMD EPYC 7R32     & \SIrange{2.8}{3.3}{\GHz} & Zen~2 (x86-64, 2019)          & DDR4 (2933\,MT/s)    \\ 
       AMD EPYC 7R13     & \SIrange{2.7}{3.7}{\GHz} & Zen~3 (x86-64, 2021)          & DDR4 (3200\,MT/s)    \\ 
       AMD EPYC 9R14     & \SIrange{3.0}{3.7}{\GHz} & Zen~4 (x86-64, 2023)          & DDR5 (4800\,MT/s)    \\ 
       Intel Xeon 8124M  & \SIrange{3.0}{3.5}{\GHz} & Skylake-SP (x86-64, 2017)     & DDR4 (2666\,MT/s)    \\ 
       Intel Xeon 8375C  & \SIrange{2.6}{3.8}{\GHz} & Ice Lake-SP (x86-64, 2021)    & DDR4 (3200\,MT/s)    \\ 
       Intel Xeon 8488C  & \SIrange{2.0}{3.8}{\GHz} & Sapphire~Ridge (x86-64, 2023) & DDR5 (4800\,MT/s)    \\ 
       \bottomrule
     \end{tabular}\restartrowcolors
   \end{minipage}
 \end{table}

\subsection{Data}%
\label{sec:data}

We use three core datasets selected to represent distinct numerical formats and
common use cases when converting floating-point numbers to strings. These
datasets, summarized in Table~\ref{tab:data}, include compactly represented
integers, high-precision floating-point numbers serialized as strings, and
synthetically generated uniformly distributed numbers.

\begin{itemize}
  \item The \emph{mesh} dataset contains vertex coordinates from a triangulated
    3D surface. Many values are small (typically in $[-1,3]$) and are
    represented with few characters, including a large proportion of exact
    integers.
  \item The \emph{canada} dataset is derived from a JSON file~\cite{langdale2019parsing}
    from the GeoJSON project, containing 64-bit floating-point numbers
    serialized as strings. These values represent geographic coordinates and
    attributes (e.g., \texttt{83.109421000000111}), and are representative of
    Geographic Information Systems (GIS) and navigation pipelines.
  \item The \emph{unit} dataset consists of uniformly generated floating-point
    numbers in the interval $[0,1)$. While synthetic, it serves as a useful
    baseline for comparison with prior work, where such distributions are
    common (e.g., to store normalized values or probabilities).
\end{itemize}
For all benchmarks, we use arrays of either 32- or 64-bit numbers. When the
original source provides only 64-bit numbers, we cast them to 32-bit prior to
benchmarking.

\begin{table}
   \caption{Dataset summary. An integer value is defined as a number exactly representable by a 64-bit signed integer. The number of digits is the minimum required for exact round-trip conversion.}%
  \label{tab:data}
  \centering
  \begin{minipage}{\textwidth}
    \centering
    \begin{tabular}{lrrrr}
      \toprule
      \multirow{2}{*}{Name} & \multirow{2}{*}{Count} & \multirow{2}{*}{Integers} & \multicolumn{2}{c}{Average digits} \\
                            &                        &                           & 32-bit & 64-bit \\
      \midrule
      mesh                  & \num{73\,019}          & \num{44\,557}             & 4.7    & 6.6    \\
      canada                & \num{111\,126}         & \num{46}                  & 7.3    & 15.3   \\
      unit                  & \num{100\,000}         & \num{0}                   & 7.5    & 16.0   \\
      \bottomrule
    \end{tabular}\restartrowcolors
  \end{minipage}
\end{table}

To better approximate real-world workloads and address the limitations of prior
benchmarks discussed in Section~\ref{sec:limitations-benchmarks}, we also
assembled a collection of additional datasets drawn from finance, astronomy,
machine learning, and meteorology~\cite{float-data}. These are summarized in
Table~\ref{tab:extra-data}. They consist entirely of floating-point values that
arise in deployed systems and public APIs, and thus complement the three core
datasets above. Specifically:
\begin{itemize}
  \item \emph{bitcoin}: daily closing prices of Bitcoin (USD), typical of
    financial APIs and market-data feeds.
  \item \emph{marine}: values from a marine-robotics inverse-kinematics
    example, representative of control and scientific-computing workloads.
  \item \emph{mobilenetv3\_large}: model weights from the
    MobileNetV3-Large ImageNet model, characteristic of machine-learning
    pipelines and neural-network parameter storage.
  \item \emph{gaia}: astrometric and photometric values from the ESA
    Gaia DR3 catalog (positions, parallaxes, fluxes), representing large-scale
    scientific data with substantial dynamic range.
  \item \emph{noaa\_global\_hourly\_2023}: surface-station telemetry
    (temperature, pressure, visibility), representative of noisy real-world
    measurement streams.
  \item \emph{noaa\_gfs}: fields extracted from NOAA GFS forecast-model
    output (temperature, humidity, wind components), representative of
    large-scale gridded scientific simulations.
\end{itemize}

\begin{table}[htbp]
  \caption{Additional real-world datasets used in some of our experiments.
  ``Integers'' counts values exactly representable as 64-bit signed integers.}
  \label{tab:extra-data}
  \centering
  \begin{tabular}{lrrrr}
    \toprule
    Name                                & Count             & Integers       & Binary type \\
    \midrule
    \texttt{bitcoin}                    & \num{943}         & \num{0}        & binary64 \\
    \texttt{marine}                     & \num{114\,950}    & \num{0}        & binary32 \\
    \texttt{mobilenetv3\_large}         & \num{5\,507\,432} & \num{0}        & binary32 \\
    \texttt{gaia}                       & \num{3\,879\,638} & \num{0}        & binary64 \\
    \texttt{noaa\_global\_hourly\_2023} & \num{1\,000\,000} & \num{428\,161} & binary32 \\
    \texttt{noaa\_gfs}                  & \num{4\,841\,536} & \num{970\,436} & binary32 \\
    \bottomrule
  \end{tabular}\restartrowcolors
\end{table}

Results in Sections~\ref{sec:none-shortest}--\ref{sec:32-bits} are reported for
the mesh, canada and unit datasets. Section~\ref{sec:extra-data} presents
additional experiments on the datasets listed in Table~\ref{tab:extra-data}.
Complete results are available in our public benchmark data repository.

\subsection{Software Implementations}%
\label{sec:software-impl}

We benchmark a selection of C and C++ libraries capable of converting IEEE
floating-point numbers to their shortest decimal string representations. Our
benchmarking code, synthetic data generators, and datasets are all publicly
available online.\footnote{\url{https://github.com/fastfloat/float_serialization_benchmark}}
The benchmarked libraries and algorithms are the following:
\begin{itemize}
  \item \emph{Grisu3 and Schubfach}: Both are evaluated using the \texttt{Drachennest}
    library.\footnote{\url{https://github.com/abolz/Drachennest}, git hash
    \texttt{e6714a3} (May 2021). Only 64-bit function is available for Grisu3.
    We exclude Grisu2, as it can produce longer-than-necessary significands.
    Drachennest also implements Dragon4, but due to concerning
    faults (see: \url{https://github.com/fastfloat/float_serialization_benchmark/pull/18}),
    we benchmark Dragon4 using a separate implementation.}
  \item \emph{Dragon4}: Benchmarked using a dedicated
    library\footnote{\url{https://github.com/lemire/Dragon4.git}, git hash
    \texttt{0ce72aa} (March 2025). Modified for portability (renamed
    \texttt{Math.h} to \texttt{DragonMath.h}). This is an implementation of
    Juckett~\cite{juckett2014} based on Burger and Dybvig's variant of Dragon4~\cite{10.1145/249069.231397},
    and is expected to be faster than a straightforward Dragon4. However, its
    64-bit implementation is not entirely correct (e.g., \texttt{5e-324} outputs a
    long string of zeros).} rather than the Drachennest version.
  \item \emph{Ryū}: Evaluated using the \texttt{Ryu}
    library.\footnote{\url{https://github.com/ulfjack/ryu}, git hash \texttt{e6714a3} (February 2024).}
  \item \emph{Dragonbox}: Benchmarked with the \texttt{Dragonbox}
    library.\footnote{\url{https://github.com/jk-jeon/dragonbox}, version 1.1.3 (June 2022).}
    The author of Dragonbox observes that the string generation they include is not
    officially part of the algorithm. They make it possible for users to provide
    their own algorithm to convert significands and exponents to strings.
  \item \emph{Google double-conversion (Grisu3-based)}: We include Google's
    \texttt{double-conversion} library.\footnote{\url{https://github.com/google/double-conversion}, version
    3.3.1 (February 2025).} We use double-conversion with the default flag.
    There are additional flags: e.g., for forcing a trailing decimal point (and
    optional zero) for integer-valued floats like "123." or for emitting '+' in
    positive exponents. We omit Google's Abseil and \texttt{snprintf}, as they
    do not guarantee shortest-string output.
  \item \emph{fmt (Dragonbox-based)}: Evaluated using the \texttt{fmt}
    library,\footnote{\url{https://github.com/fmtlib/fmt}, version
    11.1.4.} which employs a version of Dragonbox internally.
  \item \emph{SwiftDtoa}: The Swift language implementation includes a C++
    function combining ideas from Grisu2 and
    Ryū.\footnote{\url{https://github.com/swiftlang/swift.git}, git hash
    \texttt{6a862d2} (March 2025).}
  \item \emph{std::to\_chars}: C++17's standard floating-point to string function.
\end{itemize}

All algorithms above (except as noted) provide correct round-trips for both 32-
and 64-bit floats. Drachennest's Grisu3 implementation is limited to 64-bit
values. The Dragon4 implementation used is not fully correct for 64-bit; e.g.,
it mishandles subnormals such as \texttt{5e-324}.

When libraries expose flags or configuration options that alter the printed
format (e.g., forcing a trailing decimal point, controlling exponent signs, or
choosing between fixed and scientific styles), we systematically use the default
settings provided by the library.

For consistency, we focus on algorithms that generate the entire output string.
Though there are differences in the strings generated, we take these small
differences into account in our analysis (See Section~\ref{sec:perf-comp}).
Libraries like Gay's \texttt{dtoa}\footnote{Gay's dtoa:
\url{https://www.netlib.org/fp/}, retrieved January 2025.} and
teju\_jagua\footnote{Teju Jaguá: \url{https://github.com/cassioneri/teju_jagua},
git hash \texttt{e62fcfc} (March 2025).} were not benchmarked, as they compute
only the decimal significand and exponent, requiring additional
string-generation code. Though it is not difficult to implement the string
generation, such work would have an impact on the benchmarking results. We also
restrict our study to C and C++ implementations. Fair cross-language
benchmarking is outside our scope, and other languages often adopt techniques
originating in C or C++.


\subsection{None Provide the Shortest Strings}%
\label{sec:none-shortest}

A key goal in floating-point to string conversion is to produce the shortest
possible decimal string representations. Despite the centrality of string length
in practical deployments (serialization, logging, telemetry), prior benchmark
studies have never quantified end-to-end output length at scale. Our results
provide, to our knowledge, the first such characterization. We define the
number of significant digits by omitting leading and trailing zeros, so that
strings like \texttt{1.0}, \texttt{10}, and \texttt{0.1} each have exactly one
significant digit. This becomes clearer when using scientific notation:
\texttt{1E0}, \texttt{1E1}, and \texttt{1E-1}. Although tested algorithms
produce strings with the fewest digits required for exact round-trips (except
Dragon4 for 64-bit numbers due to the aforementioned bug in the available
implementation), none consistently generate the shortest strings in terms of
total character length.\footnote{We assess shortest-string behavior by
cross-comparing output lengths across algorithms and against the valid decimal
representation already present in the dataset. A shorter dataset string
implies that none of the tested algorithms produced a minimal-length result.}
For instance, the C++17 standard library's \texttt{std::to\_chars} renders the
number $0.00011$ as \texttt{0.00011} (7 characters), while the shorter
scientific form \texttt{1.1e-4} (6 characters) is possible. Similarly, it
outputs \texttt{12300} as \texttt{1.23e+04} (8 characters) rather than the
shorter \texttt{1.23e4} (6 characters). Such longer outputs result from
formatting rules inherited from the C standard, particularly regarding
scientific notation (\texttt{\%e}):

\begin{quote}
  A double argument representing a floating-point number is converted in the style \texttt{[-]d.ddde±dd}. \texttt{[\ldots]},
  The exponent always contains at least two digits, and only as many more digits.
\end{quote}

These rules mandate a positive exponent sign and at least two digits for the
exponent, following historical precedent. When converting floating-point
numbers, \texttt{std::to\_chars} chooses the shortest notation between
fixed-point (\texttt{\%f}) and scientific (\texttt{\%e}), favoring fixed-point
notation if lengths are equal.\footnote{For example, the 32-bit value \texttt{4.27819e+09} can be equally represented by the strings \texttt{4.27819e+09} (scientific), \texttt{4278190080}, or \texttt{4278190000} (both fixed-point); all have the same string length (10 characters). The C++ standard requires choosing fixed-point notation in case of a tie. Among possible fixed-point outputs, the standard then mandates selection of the string numerically closest to the exact value. Here, \texttt{4278190080} (with 9 significant digits) is chosen over \texttt{4278190000} (6 significant digits), even though it has more digits, since significant digits are not considered, only string length and numerical proximity.}
Similar constraints affect libraries such as fmt, Grisu3, Schubfach,
double\_conversion, and swiftDtoa. In contrast, Ryū, Dragonbox, and Grisu-Exact
do not strictly follow these rules but frequently prefer scientific notation
even when longer (e.g., printing \texttt{0.1} as \texttt{1E-1}).

Table~\ref{tab:avglength} summarizes the average string length in characters
across our three core datasets. For 32-bit numbers, algorithms like Dragon4,
double\_conversion, fmt, and \texttt{std::to\_chars} consistently produce the
shortest average lengths across the mesh (approximately \SI{5.9}{\characters}),
canada (\SI{8.8}{\characters}), and unit (approximately \SI{9.6}{\characters})
datasets. Conversely, Grisu-Exact, Ryū, and Dragonbox consistently yield longer
strings (\SIrange{7.7}{8.0}{\characters} for mesh; \SI{10.8}{\characters} for
canada; \SI{11.5}{\characters} for unit).

\begin{table}
  \caption{Average number of characters. We use the libstdc++ \texttt{std::to\_chars} implementation.}%
  \label{tab:avglength}
  \centering
  \begin{tabular}{lrrrrrr}
    \toprule
    \multicolumn{1}{c}{Name}& \multicolumn{2}{c|}{mesh} & \multicolumn{2}{c|}{canada} & \multicolumn{2}{c}{unit} \\
                            & 32-bit  & 64-bit          & 32-bit   & 64-bit           & 32-bit & 64-bit \\ \midrule
    Dragon4                 & 5.863   & 7.589           & 8.823    & 16.800           & 9.628  & 18.273 \\
    fmt                     & 5.913   & 7.589           & 8.823    & 16.800           & 9.627  & 18.272 \\
    Grisu3                  & --      & 7.589           & --       & 16.800           & --     & 18.273 \\
    Grisu-Exact             & 7.697   & 9.571           & 10.824   & 18.800           & 11.515 & 20.160 \\
    Schubfach               & 5.863   & 7.589           & 8.823    & 16.800           & 9.628  & 18.273 \\
    Dragonbox               & 8.005   & 9.879           & 10.824   & 18.800           & 11.515 & 20.160 \\
    Ryū                     & 8.005   & 9.879           & 10.824   & 18.800           & 11.515 & 20.160 \\
    double\_conversion      & 5.863   & 7.737           & 8.823    & 16.800           & 9.627  & 18.272 \\
    swiftDtoa               & 7.034   & 8.810           & 8.825    & 16.801           & 9.627  & 18.272 \\
    \texttt{std::to\_chars} & 5.863   & 7.589           & 8.823    & 16.800           & 9.627  & 18.272 \\ \midrule
    Shortest                & 4.537   & 6.263           & 8.823    & 16.800           & 9.626  & 18.268 \\
    \bottomrule
  \end{tabular}\restartrowcolors
\end{table}

Similar trends emerge for 64-bit numbers: Dragon4, fmt, Grisu3, Schubfach, and
\texttt{std::to\_chars} produce shorter average lengths (7.6~for mesh, 16.8~for
canada, and around 18.3~for unit), closely followed by double\_conversion
and swiftDtoa. Grisu-Exact, Ryū, and Dragonbox consistently produce longer
strings (approximately 9.9 for mesh, 18.8 for canada, and 20.16 for unit).

The \emph{Shortest} row in Table~\ref{tab:avglength} illustrates the optimal
achievable lengths. Notably, on the mesh dataset, \texttt{std::to\_chars}
produces strings that are significantly longer than optimal. For example, its
outputs are roughly 30\% longer than the true minimum for 32-bit numbers and
about 20\% longer for 64-bit numbers. These discrepancies show that differences
in total character length between implementations are substantial, underscoring
the importance of evaluating both digit count and final string length when
benchmarking conversion algorithms.

This behavior is not a correctness issue, but rather a deliberate design
decision shared by nearly all modern algorithms, which are engineered to
minimize the number of significant digits of the decimal significand---not
the total character length of the printed string. Formatting decisions (fixed
vs.\ scientific notation, exponent width, trailing zeros) are delegated to the
caller or the standard library. Our measurements show that this separation can
yield \emph{unexpectedly large} gaps between the shortest valid significand and
the shortest possible printed representation---sometimes exceeding 30\%. This
highlights a practical yet previously overlooked distinction between
\emph{shortest significand} and \emph{shortest string}.

\subsection{Performance Comparison: Schubfach and Dragonbox are faster}%
\label{sec:perf-comp}

We compiled all functions in release mode using the default CMake parameters
(\texttt{-O3 -DNDEBUG}). When possible (e.g., on x86-64 systems), we enabled
hardware-specific optimizations (\texttt{-march=native}). CPU performance
counters were used to record the number of completed instructions and CPU
cycles, in addition to wall-clock time, to better capture microarchitectural effects.

Tables~\ref{tab:appleresults} and~\ref{tab:ryzen9900xresults} present detailed
benchmarking results for 64-bit float-to-string conversion on Apple M4 Max and
AMD Ryzen~9900X processors. We report three key metrics: nanoseconds per float
(ns/f), instructions per float (ins/f), and instructions per cycle (ins/c), to
capture both algorithmic efficiency and hardware utilization. While previous
work has occasionally reported wall-clock timing, instruction-level metrics for
float-to-string conversion (in particular ins/f and ins/c) have not been
systematically analyzed in the literature. We use these to distinguish intrinsic
algorithmic cost (ins/f) from microarchitectural utilization (ins/c), exposing
differences invisible to timing-only evaluations.

To assess measurement stability, each benchmark was repeated 100 times. On
dedicated hardware (Apple M4 Max, AMD Ryzen~9900X), timing variability was
consistently low (median $0.7\%$, max $2.7\%$). In contrast, on cloud instances,
a handful of runs showed higher variability (median $0.9\%$, max $6.9\%$). These
outliers likely reflect intermittent resource contention or infrastructure noise
inherent to virtualized environments, but appear to affect all algorithms
similarly. Cycles per float (\texttt{c/f}) tracked timing variability closely,
with a median variation of $0.4\%$, though rare outliers again reached $6\%$. In
all cases, instructions per float (\texttt{ins/f}) remained completely
deterministic (0.0\% variability), confirming that the executed instruction path
is unaffected by runtime fluctuations. Observed performance jitter therefore
arises solely from external factors impacting timing and cycle counts, rather
than any non-determinism in the algorithms themselves.

\begin{table}[htpb]
  \centering
  \caption{Apple M4 Max results (Apple/LLVM 17, 64-bit floats)}%
  \label{tab:appleresults}
  \begin{tabular}{lccccccccc}
    \toprule
    \multicolumn{1}{c}{Name}& \multicolumn{3}{c|}{mesh}  & \multicolumn{3}{c|}{canada} & \multicolumn{3}{c}{unit}   \\
                            & {ns/f} & {ins/f} & {ins/c} & {ns/f} & {ins/f} & {ins/c}  & {ns/f} & {ins/f} & {ins/c} \\ \midrule
    Dragon4                 & 69     & 1500    & 5.3     & 150    & 3000    & 4.8      & 170    & 3300    & 4.6     \\
    fmt                     & 22     & 530     & 5.4     & 29     & 640     & 5.0      & 30     & 510     & 3.8     \\
    Grisu3                  & 10     & 260     & 5.6     & 24     & 440     & 4.2      & 26     & 470     & 4.0     \\
    Grisu-Exact             & 11     & 320     & 6.3     & 15     & 340     & 5.1      & 18     & 340     & 4.2     \\
    Schubfach               & 7.2    & 210     & 6.4     & 12     & 310     & 5.9      & 14     & 290     & 4.7     \\
    Dragonbox               & 7.7    & 220     & 6.6     & 9.5    & 240     & 5.6      & 12     & 230     & 4.2     \\
    Ryū                     & 9.9    & 270     & 6.0     & 12     & 330     & 6.3      & 13     & 310     & 5.4     \\
    double\_conversion      & 26     & 640     & 5.5     & 42     & 910     & 5.1      & 43     & 880     & 4.8     \\
    swiftDtoa               & 14     & 390     & 6.0     & 16     & 360     & 5.1      & 20     & 390     & 4.4     \\
    \texttt{std::to\_chars} & 13     & 350     & 5.8     & 15     & 440     & 6.6      & 16     & 410     & 5.6     \\
    \bottomrule
  \end{tabular}\restartrowcolors
\end{table}

\begin{table}[htpb]
  \centering
  \caption{AMD Ryzen 9900X results (g++15, 64-bit floats)}%
  \label{tab:ryzen9900xresults}
  \begin{tabular}{lccccccccc}
    \toprule
    \multicolumn{1}{c}{Name}& \multicolumn{3}{c|}{mesh}  & \multicolumn{3}{c|}{canada} & \multicolumn{3}{c}{unit}   \\
                            & {ns/f} & {ins/f} & {ins/c} & {ns/f} & {ins/f} & {ins/c}  & {ns/f} & {ins/f} & {ins/c} \\ \midrule
    Dragon4                 & 82     & 2300    & 5.0     & 170    & 4700    & 5.2      & 190    & 5000    & 4.9     \\
    fmt                     & 30     & 570     & 3.5     & 40     & 840     & 3.8      & 35     & 560     & 2.9     \\
    Grisu3                  & 12     & 290     & 4.5     & 29     & 630     & 4.0      & 26     & 510     & 3.6     \\
    Grisu-Exact             & 18     & 370     & 3.7     & 24     & 520     & 3.9      & 21     & 370     & 3.1     \\
    Schubfach               & 9.9    & 250     & 4.5     & 24     & 490     & 3.7      & 19     & 320     & 3.0     \\
    Dragonbox               & 11     & 260     & 4.3     & 18     & 410     & 4.1      & 15     & 240     & 3.0     \\
    Ryū                     & 14     & 320     & 4.3     & 24     & 580     & 4.3      & 20     & 400     & 3.5     \\
    double\_conversion      & 27     & 610     & 4.0     & 45     & 1000    & 4.0      & 39     & 810     & 3.7     \\
    swiftDtoa               & 23     & 490     & 3.8     & 28     & 590     & 3.8      & 27     & 440     & 3.0     \\
    \texttt{std::to\_chars} & 18     & 490     & 4.8     & 30     & 780     & 4.8      & 25     & 600     & 4.3     \\
    \bottomrule
  \end{tabular}\restartrowcolors
\end{table}

Across all datasets on the Apple M4 Max processor
(Table~\ref{tab:appleresults}), Dragonbox and Schubfach consistently achieve the
fastest performance, with runtime (ns/f) measures ranging from \num{7.2} to
\num{14}. Schubfach notably achieves the lowest runtime on the mesh dataset
(\SI{7.2}{\nano\second\per\float}), while Dragonbox leads on the canada dataset
(\SI{9.5}{\nano\second\per\float}). Both algorithms also show minimal instruction
counts (\SIrange{210}{310}{\instruction\per\float}), reflecting efficient
implementations. Ryū closely follows, with runtimes ranging from
\SIrange{9.9}{13}{\nano\second\per\float} and instruction counts between \num{270}
and \num{330} ins/f. On the Apple M4 Max, Dragon4 is consistently the slowest,
with ns/f measures between \num{69} (mesh) and \num{170} (unit), and significantly
higher instruction counts (\SIrange{1500}{3300}{\instruction/\float}).
The double\_conversion and fmt functions exhibit moderate performance
(\SIrange{22}{43}{\nano\second\per\float}), substantially faster than Dragon4 but
slower than Dragonbox and Schubfach. The remaining algorithms---Grisu-Exact,
Grisu3, swiftDtoa, and \texttt{std::to\_chars}---occupy the intermediate
performance range (\SIrange{10}{26}{\nano\second\per\float}). Performance trends
are similar on the AMD Ryzen~9900X processor
(Table~\ref{tab:ryzen9900xresults}). Schubfach and Dragonbox again perform best
across datasets (e.g., Schubfach at \SI{9.9}{\nano\second\per\float}, Dragonbox at
\SI{11}{\nano\second\per\float} for mesh). Dragon4 remains the slowest, with even
higher instruction counts (up to \num{5000} ins/f on unit) and longer runtimes
(up to \SI{190}{\nano\second\per\float}). The relative ranking of other algorithms
(\texttt{std::to\_chars}, double\_conversion, fmt, etc.) remains largely
consistent with the Apple M4 Max results.

Differences in algorithm outputs---particularly string lengths---must be
considered when interpreting these results. Dragonbox and Ryū, for instance,
often produce longer strings than Schubfach and \texttt{std::to\_chars},
potentially influencing relative runtimes. Nevertheless, large performance gaps
persist even among algorithms generating comparable string lengths (e.g.,
Schubfach vs.\ fmt), underscoring genuine algorithmic efficiency differences.

A consistent trend across both the Apple M4 Max and AMD Ryzen~9900X results is
that all algorithms perform fastest on the mesh dataset, are slower on the
canada dataset, and slowest on the unit dataset (as indicated by the ns/f column
in Tables~\ref{tab:appleresults} and~\ref{tab:ryzen9900xresults}). This ranking
reflects the differences in output string lengths reported in
Table~\ref{tab:avglength}, with the unit dataset producing the longest outputs.
While it is tempting to attribute the slowdown solely to the increased cost of
formatting longer strings, the instructions-per-cycle (ins/c) results reveal an
additional effect: all algorithms achieve lower ins/c on the unit dataset
compared to mesh, indicating reduced pipeline throughput.

To further investigate, we profiled the execution of the \texttt{std::to\_chars}
algorithm on the Ryzen~9900X. On the unit dataset, 28\% of executed
instructions and 34\% of cycles occur within the string formatting routine.
For comparison, on the mesh dataset, these figures are notably lower: 19\% of
instructions and 21\% of cycles. This increase confirms that longer outputs
entail a greater share of processing within the formatting function, yet a
substantial portion of instructions and cycles remains attributable to other
algorithm components. Furthermore, the higher ratio of cycles to
instructions spent in string formatting on the unit dataset suggests
microarchitectural bottlenecks, such as increased memory latency or branch
misprediction. Overall, these results indicate that the performance degradation
on the unit dataset arises from both increased output costs and diminished
execution efficiency within the processor pipeline.

\begin{figure}[p]
  \centering
  \includegraphics[width=\textwidth]{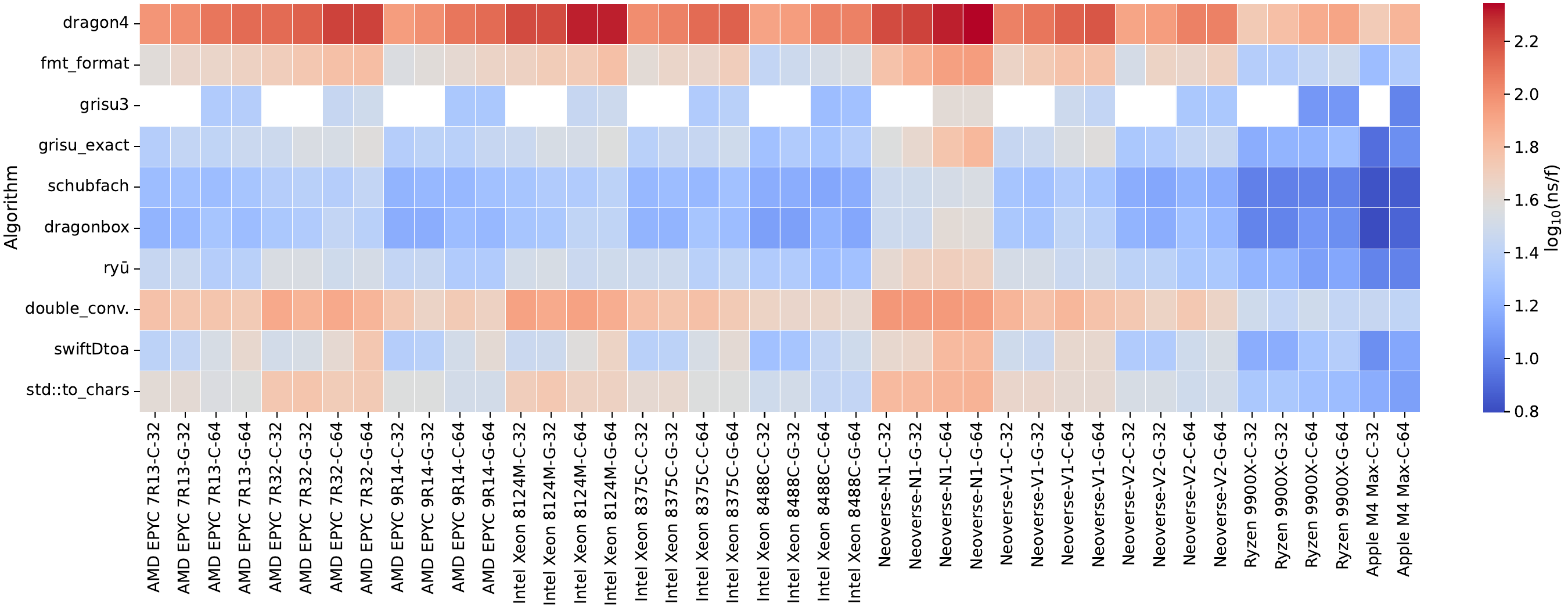}
  \caption{Algorithm performance (log$_{10}$ ns/f) across CPUs, compilers, and widths (mesh dataset)}%
  \label{fig:heatmap_mesh}
\end{figure}

\begin{figure}[p]
  \centering
  \includegraphics[width=\textwidth]{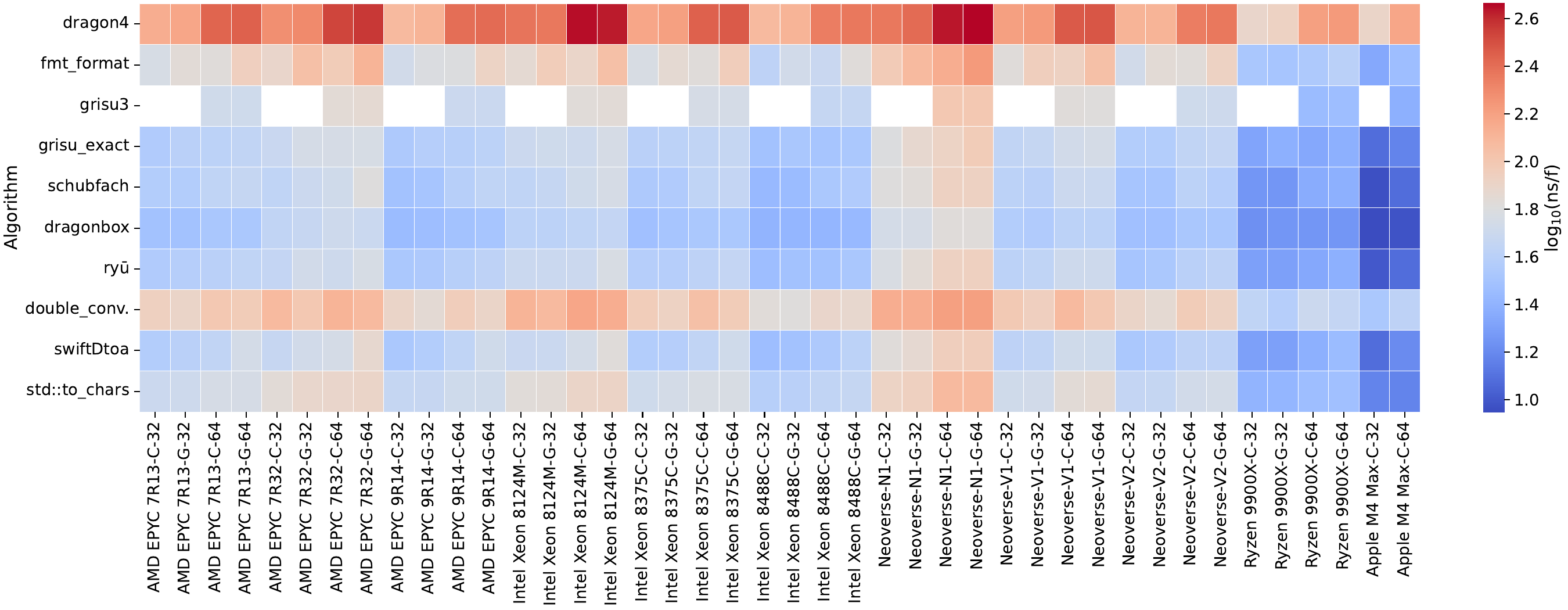}
  \caption{Algorithm performance (log$_{10}$ ns/f) across CPUs, compilers, and widths (canada dataset)}%
  \label{fig:heatmap_canada}
\end{figure}

\begin{figure}[p]
  \centering
  \includegraphics[width=\textwidth]{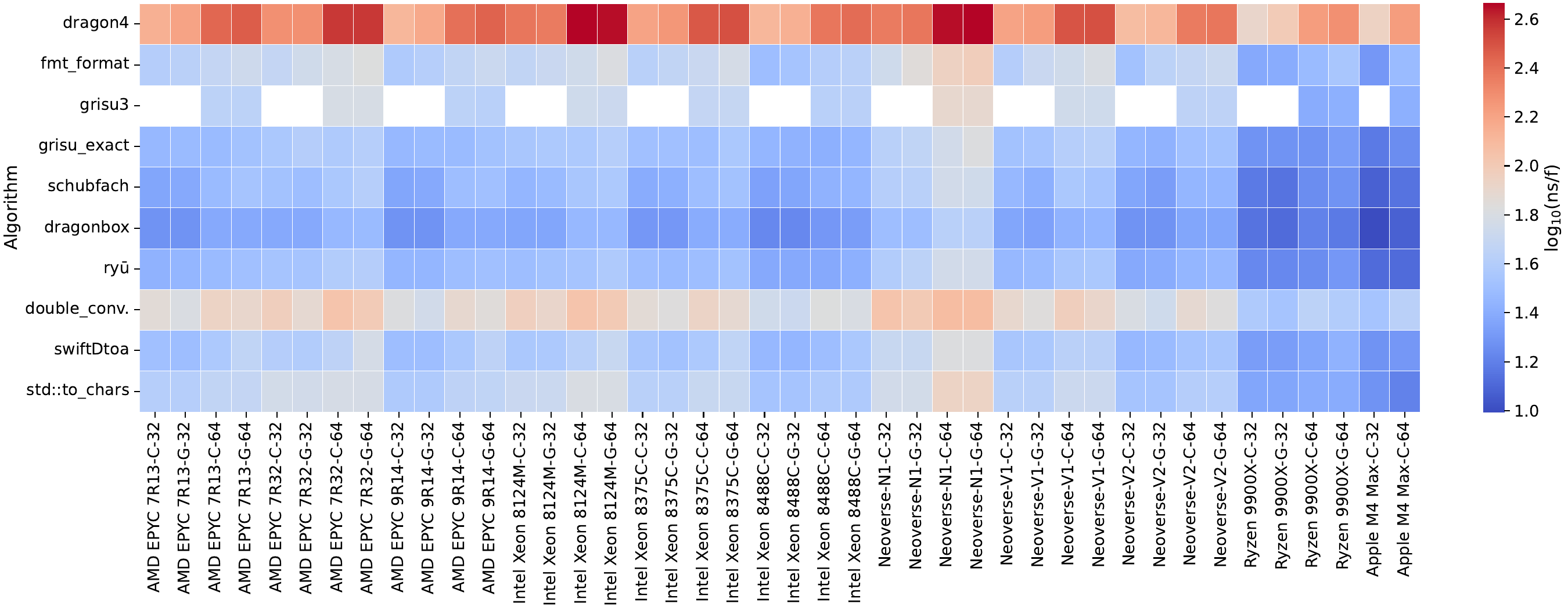}
  \caption{Algorithm performance (log$_{10}$ ns/f) across CPUs, compilers, and widths (unit dataset)}%
  \label{fig:heatmap_unit}
\end{figure}

\begin{figure}
  \centering
  \includegraphics[width=0.83\textwidth]{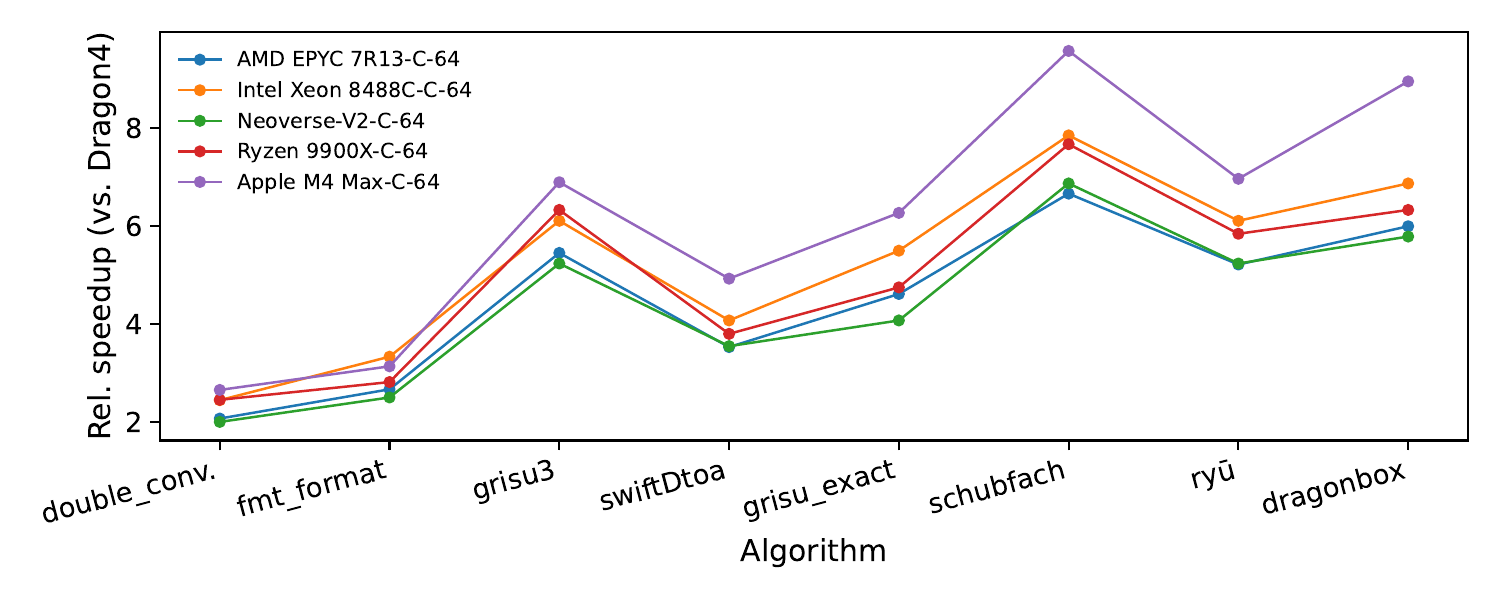}
  \caption{Relative speedup (vs.\ \texttt{dragon4}) for selected CPUs (mesh dataset)}%
  \label{fig:relative_performance_mesh}
\end{figure}

\begin{figure}
  \centering
  \includegraphics[width=0.83\textwidth]{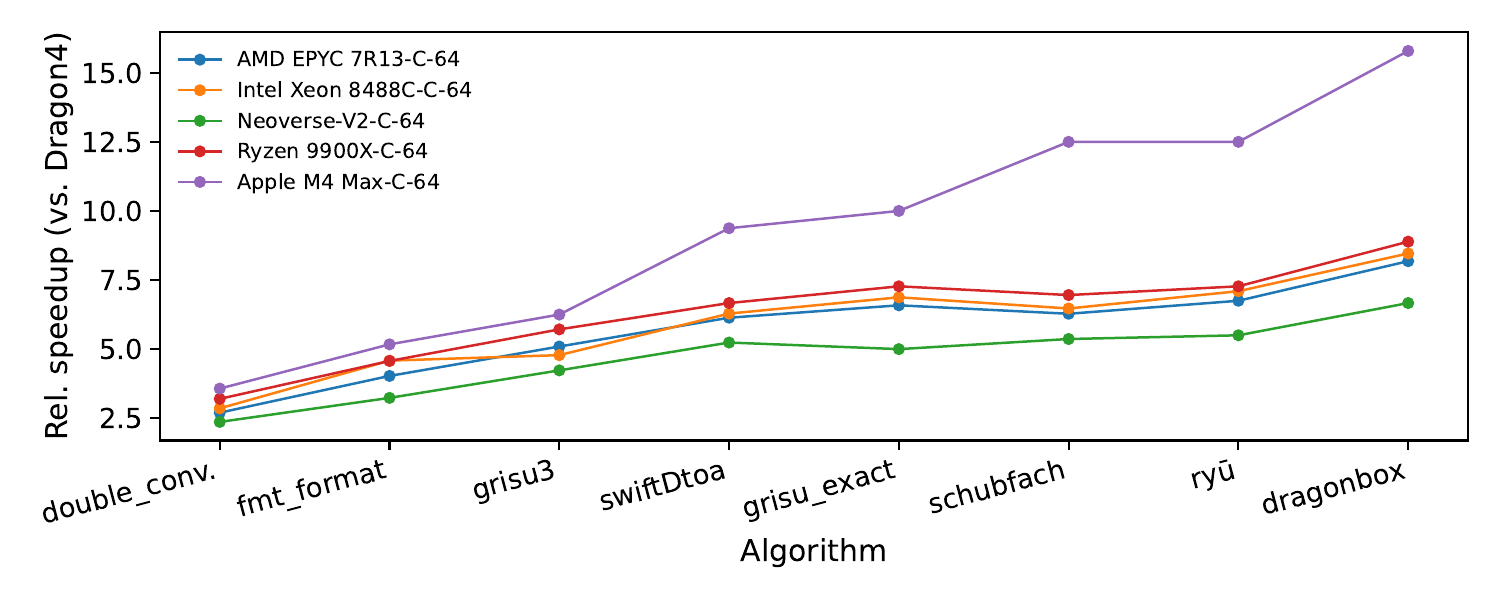}
  \caption{Relative speedup (vs.\ \texttt{dragon4}) for selected CPUs (canada dataset)}%
  \label{fig:relative_performance_canada}
\end{figure}

\begin{figure}
  \centering
  \includegraphics[width=0.83\textwidth]{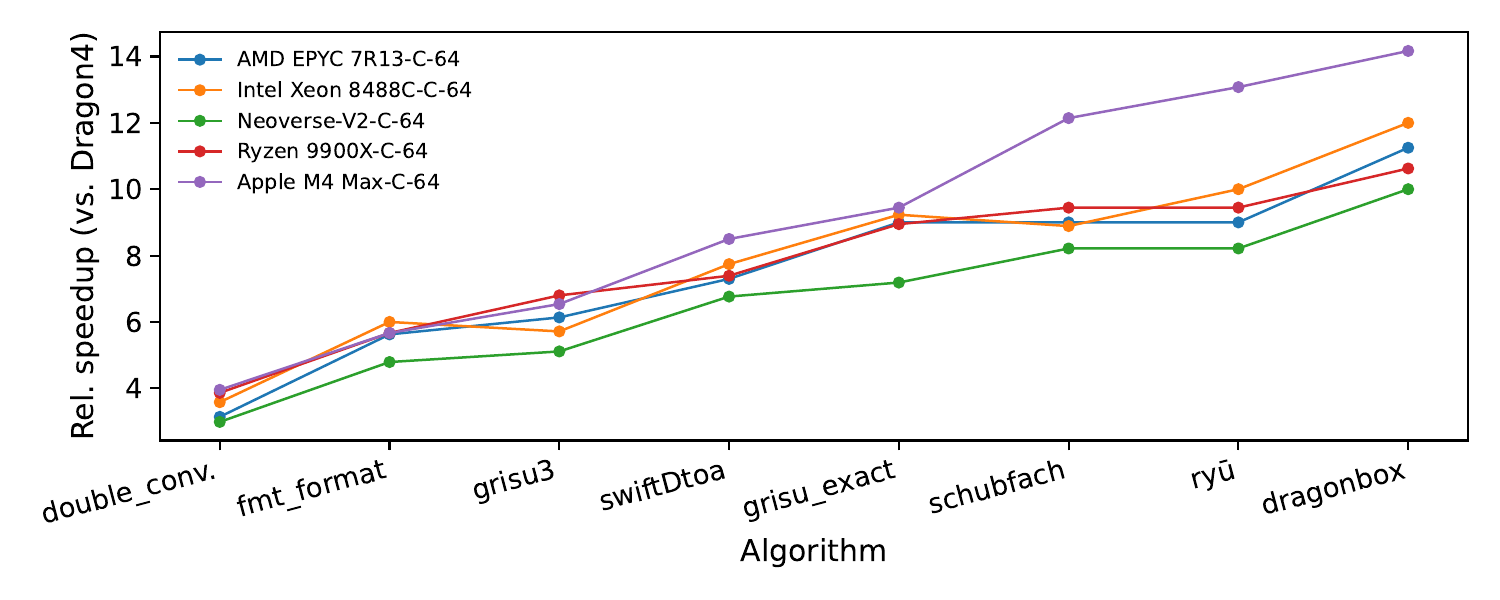}
  \caption{Relative speedup (vs.\ \texttt{dragon4}) for selected CPUs (unit dataset)}%
  \label{fig:relative_performance_unit}
\end{figure}

Figures~\ref{fig:heatmap_mesh},~\ref{fig:heatmap_canada},~and~\ref{fig:heatmap_unit}
visualize algorithmic performance (log$_{10}$ ns/f) across datasets, CPUs, compilers, and floating-point widths.\footnote{Raw benchmark data for these visualizations are available on the paper's website at \url{https://www.jaelgareau.com/en/publication/gareau_lemire-spe25}.}
In these heatmaps, dark blue regions correspond to the fastest execution speeds,
transitioning through light blue and light red to dark red, which indicates the
slowest performance. This gradient clearly highlights the influence of compiler
choice, CPU architecture, and numerical width (32-bit vs.\ 64-bit). Several
notable insights emerge:
\begin{itemize}
  \item \emph{Compiler choice affects algorithmic performance}. On the mesh dataset,
    Schubfach often runs faster when compiled with clang++ (with libc++), while
    Dragonbox frequently benefits from g++ (with libstdc++). For example, on the
    Xeon 8488C CPU, Schubfach was approximately 7\% faster with clang++, whereas
    Dragonbox was 12.5\% faster with g++. This effect is observed on the
    majority of CPUs in our study, though the exact magnitude of the compiler
    advantage varies across architectures and algorithms.
  \item \emph{CPU architecture influences overall performance}.
    The Neoverse~N1 (Graviton~2), with three arithmetic units but only one capable of
    executing multiplications, consistently exhibits slower performance across
    algorithms. In contrast, recent high-end processors like AMD~Zen~5 feature
    six arithmetic units, three of which can perform multiplications in
    parallel. This greater arithmetic parallelism likely contributes to Apple’s
    M4 Max and similar CPUs typically ranking among the fastest in our tests.
    This interpretation is further supported by our measurements: the mean
    instructions-per-cycle (ins/c) across all algorithms and datasets is 3.8 on
    the Ryzen~9900X, compared to just 2.6 on the Graviton2. Such differences
    in ins/c reflect the ability of more advanced CPUs to execute a greater
    number of instructions in parallel.
  \item \emph{Algorithmic choice dominates relative performance.} While CPU
    architecture and compiler matter, the number of instructions required by
    each algorithm varies far more, making algorithm selection the single
    largest determinant of speed. For example, on the Ryzen~9900X and the unit
    dataset, ins/f values ranged from 5000 (dragon4) to 240 (dragonbox), whereas
    ins/c values across all CPUs and algorithms stayed between 2.3 and 5.6.
To further support our observation,
figures~\ref{fig:relative_performance_mesh},~\ref{fig:relative_performance_canada}~and~\ref{fig:relative_performance_unit}
show the relative performance of the algorithms compared to Dragon4 on selected
CPUs. The Apple CPU achieves relatively high performance with Schubfach, Ryū
and Dragonbox compared to Dragon4. In other words, it benefits more from a
switch to the more recent algorithms than the other selected CPUs. The
Neoverse~V2 processor has slightly lower relative performance than the other
selected CPUs. Yet the curves are visibly correlated: on the canada and unit
dataset, Dragonbox gives the best results while on the mesh dataset, Schubfach
is slightly superior to Dragonbox. The other processors have similar ratios
\end{itemize}

\subsection{Advanced CPU instructions are not exploited}%
\label{sec:advanced-instructions}

We also investigated whether recent float-to-string algorithms are able to
leverage advanced instructions available on modern CPUs, such as fused
multiply-add (FMA) and vectorization (SIMD). On x86-64, CPUs are grouped into
distinct architectural levels (x86-64-v1, v2, v3, and v4), each introducing new instruction sets and
capabilities.\footnote{See \url{https://en.wikipedia.org/wiki/X86-64\#Microarchitecture_levels}.}
The x86-64-v1 level is the baseline introduced by AMD in 2003, and it includes
foundational 64-bit capabilities like CMOV, SSE, and SSE2, compatible with early
processors like AMD~K8 and Intel Prescott. The x86-64-v2 level, defined in 2020
by AMD, Intel, Red Hat, and SUSE, adds instructions such as SSE3, SSE4.1,
SSE4.2, and POPCNT, aligning with processors from around 2008--2011, like Intel
Nehalem. The x86-64-v3 level introduces AVX, AVX2, FMA, and MOVBE, targeting
CPUs from 2013--2015, such as Intel Haswell. The x86-64-v4 level incorporates
AVX-512, doubling vector instruction width to 512 bits, and is supported by
newer CPUs like AMD Zen~4 and Zen~5.

Table~\ref{tab:architecture_levels} summarizes the performance of Schubfach and
Dragonbox across all architectural levels on the Ryzen~9900X (Zen~5). The
results show that enabling newer instructions provides, at best, marginal
improvements in performance. In one instance (Dragonbox on the unit data), the
version of our software compiled for the most advanced (x86-64-v4) instructions
required slightly more instructions. In another (Schubfach on the canada data),
targeting x86-64-v4 saved a small number of instructions. This suggests that the
Schubfach and Dragonbox implementations do not benefit from advanced
instructions on x86-64 processors.

\begin{table}
  \centering
  \caption{Schubfach vs. Dragonbox across architectural levels (g++15, 64-bit floats, Ryzen~9900X).}
  \label{tab:architecture_levels}
  \begin{tabular}{llccccccccc}
    \toprule
    \multicolumn{1}{c}{Level} & \multicolumn{1}{c}{Name} & \multicolumn{3}{c|}{mesh}  & \multicolumn{3}{c|}{canada} & \multicolumn{3}{c}{unit} \\
                              &                          & {ns/f} & {ins/f} & {ins/c} & {ns/f} & {ins/f} & {ins/c}  & {ns/f} & {ins/f} & {ins/c} \\ \midrule
    \rowcolor{black!10}
    x86-64-v1           & schubfach & 9.8 & 250 & 4.6 & 23 & 490 & 3.9 & 18 & 310 & 3.2  \\
    \rowcolor{black!10}
                        & dragonbox & 11  & 260 & 4.2 & 18 & 410 & 4.0 & 15 & 240 & 3.0  \\
    \rowcolor{white}
    x86-64-v2 & schubfach & 9.7 & 250 & 4.7  & 23 & 490 & 3.9  & 18 & 310 & 3.2  \\
    \rowcolor{white}
              & dragonbox & 11 & 260 & 4.2  & 18 & 410 & 4.1  & 15 & 240 & 2.9  \\
    \rowcolor{black!10}
    x86-64-v3 & schubfach & 9.7 & 250 & 4.7  & 23 & 490 & 3.9  & 18 & 310 & 3.2  \\
    \rowcolor{black!10}
              & dragonbox & 11 & 260 & 4.2  & 18 & 410 & 4.1  & 15 & 240 & 2.9  \\
    \rowcolor{white}
    x86-64-v4 & schubfach & 9.8 & 240 & 4.6  & 24 & 480 & 3.6  & 19 & 310 & 2.9  \\
    \rowcolor{white}
              & dragonbox & 11 & 260 & 4.2  & 18 & 410 & 4.1  & 15 & 250 & 3.0  \\
    \bottomrule
  \end{tabular}\restartrowcolors
\end{table}

\subsection{Converting 32-bit numbers may be faster?}%
\label{sec:32-bits}

When converting floating-point numbers to strings, the output is typically
shorter for 32-bit than for 64-bit values: at most nine digits suffice for
32-bit numbers, while up to seventeen are needed for 64-bit numbers. We might
therefore expect that 32-bit conversions are generally faster, since less work
is required for formatting and string generation.

Figure~\ref{fig:compare32to64} confirms this expectation on the Apple M4 Max
processor. The fastest implementations convert over \SI{96}{\Mfps} for 32-bit
floats (Schubfach: \SI{109}{\Mfps}, Dragonbox: \SI{112}{\Mfps}, Ryū:
\SI{96}{\Mfps}), while their 64-bit performance is typically lower (Schubfach:
\SI{83}{\Mfps}, Dragonbox: \SI{106}{\Mfps}, Ryū: \SI{83}{\Mfps}). The difference
is especially pronounced for the slowest algorithm: Dragon4 processes only
\SI{12}{\Mfps} for 32-bit versus \SI{7}{\Mfps} for 64-bit. Interestingly, some
algorithms---particularly \texttt{std::to\_chars} and Dragonbox---show
little or no difference between 32- and 64-bit speeds (e.g.,
\texttt{std::to\_chars} achieves \SI{66}{\Mfps} for both widths; Dragonbox
reaches \SI{112}{\Mfps} for 32-bit and \SI{106}{\Mfps} for 64-bit). This
suggests that in certain libraries, the conversion routine is dominated by fixed
overhead, or that the core bottleneck is not string length but the underlying
algorithm.

Another notable point is that \emph{character throughput} (total characters
produced per second) is higher for 64-bit numbers, since their decimal
representations are, on average, nearly twice as long. Thus, while more numbers
can be processed per second in the 32-bit case, more textual data can be
produced per second in the 64-bit case.

\definecolor{col1}{RGB}{23, 37, 84}
\definecolor{col2}{RGB}{107, 114, 128}
\begin{figure}
  \centering
  \begin{tikzpicture}
    \begin{axis}[
      height=6.9cm,
      ybar,
      bar width=0.25cm,
      enlarge x limits=0.9,
      legend style={at={(0.5,1.)}, anchor=south, legend columns=-1, draw=none},
      ylabel={millions of floats per second},
      symbolic x coords={Dragon4, fmt, swiftDtoa, std::to\_chars, Ryū, Schubfach, Dragonbox},
      xtick=data,
      x tick label style={rotate=90, anchor=north east},
      ymin=0,
      nodes near coords,
      nodes near coords align={vertical},
      every node near coord/.append style={font=\footnotesize},
      axis line style={draw=none},
      axis x line=bottom,
      ymajorgrids,
      yticklabel shift={0.2cm},
      clip=false,
      major grid style={gray!50}
    ]

      \addplot+[ybar, color=col1,fill=col1,bar shift=0.05cm] coordinates {
          (Dragon4, 12)
          (fmt, 45)
          (swiftDtoa, 82)
          (std::to\_chars, 66)
          (Ryū, 96)
          (Schubfach, 109)
          (Dragonbox, 112)
        };

      \addplot+[ybar, fill=col2, color=col2, bar shift=0.40cm,draw=none,] coordinates {
          (Dragon4, 7)
          (fmt, 34)
          (swiftDtoa, 62)
          (std::to\_chars, 66)
          (Ryū, 83)
          (Schubfach, 83)
          (Dragonbox, 106)
        };
      \legend{32-bit, 64-bit}
    \end{axis}
  \end{tikzpicture}
  \caption{Throughput (Mfloat/s) on Apple M4 Max (canada dataset, 32-/64-bit)\label{fig:compare32to64}}
\end{figure}
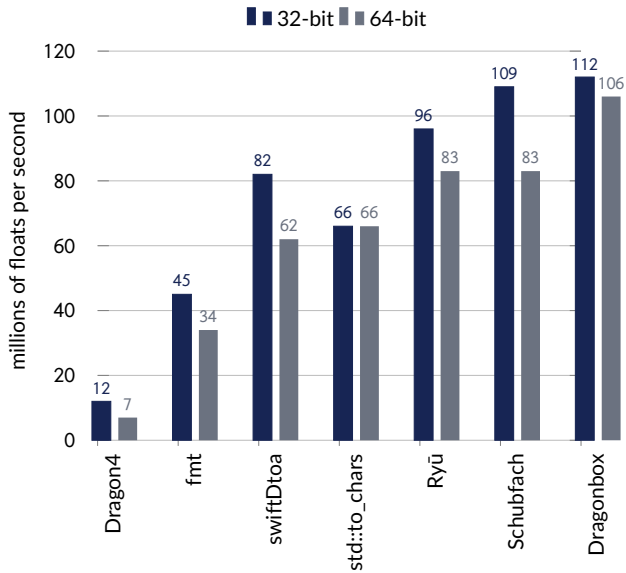

\subsection{Additional real-world datasets}
\label{sec:extra-data}

To verify that our conclusions are not an artifact of the three core datasets,
we also benchmarked all implementations on the additional real-world datasets
listed in Table~\ref{tab:extra-data}. For space reasons, we report here only a
condensed view of the results. We ran the full benchmark suite on these datasets
on both the Apple M4 Max and AMD Ryzen 9 9900X processors described in
Table~\ref{tab:test-cpus}, using the same compiler configurations as in
Section~\ref{sec:systems}.

Table~\ref{tab:additional_summary} reports detailed results for the three
implementations we use in this section---Dragonbox, Schubfach, and Ryū---which
are representative of the performance trends observed across all algorithms in
the full benchmark. The absolute numbers vary across datasets and architectures,
but the relative ordering is  consistent with our
findings for mesh, canada, and unit: Dragonbox and Schubfach are typically the
fastest, while Ryū is consistently slower by a moderate but systematic margin.

Several patterns mirror those observed earlier. First, algorithms that emit
longer strings---particularly Ryū and Dragonbox---show slightly higher
instruction counts on high-dynamic-range datasets, such as \texttt{bitcoin} or
\texttt{gaia}. Second, datasets dominated by exact integers or very small
magnitudes (e.g., \texttt{noaa\_global\_hourly\_2023} and \texttt{marine\_ik})
yield the shortest strings and therefore the fastest conversions among the
real-world datasets. Datasets that are still relatively small but contain fewer
integers (e.g., \texttt{noaa\_gfs}) produce moderately longer strings and
slightly slower performance, though they are still faster than
high-dynamic-range sets.

Crucially, none of the additional datasets reveals outliers that contradict our
earlier conclusions. Across all workloads, Dragonbox and Schubfach remain among
the fastest implementations, Ryū is competitive but consistently slower, and
the performance gaps fall squarely within the ranges observed on the core
datasets.

String-length behavior generalizes just as cleanly. None of the implementations
reaches the theoretically shortest possible strings, and the excess is again
frequently in the 20--30\% range, particularly on datasets with large exponents
such as \texttt{bitcoin} and \texttt{gaia}. This reinforces our earlier finding
that minimizing the number of significant digits is insufficient to guarantee
shortest-string output, and that real-world workloads tend to amplify these
effects.

Overall, these results support the robustness of our conclusions across a broad
range of practical numerical domains, including finance, astronomy, machine
learning, and meteorology. Detailed per-dataset results for all
algorithms are available in our public benchmark repository.

\begin{table*}[htbp]
  \centering
  \caption{Performance (ns/f, ins/f, ins/c) of Dragonbox, Ryū, and Schubfach on
  six additional real-world datasets. Results are shown for two CPUs (Ryzen 9 9900X,
  Apple M4 Max) and exhibit the same ranking and trends seen on the core datasets.}%
  \label{tab:additional_summary}
  \begin{tabular}{llrrrrrr}
    \toprule
    Dataset & Algorithm & \multicolumn{3}{c}{AMD Ryzen 9 9900X} & \multicolumn{3}{c}{Apple M4 Max} \\
                                     &                    & ns/f  & ins/f & ins/c & ns/f  & ins/f & ins/c \\
    \midrule
    \rowcolor{black!10}
    bitcoin (f64)                    & dragonbox          & 12.0  & 290   & 4.3   & 8.6   & 270   & 6.4 \\
    \rowcolor{black!10}
                                     & ryu                & 23.0  & 530   & 4.1   & 13.0  & 420   & 6.6 \\
    \rowcolor{black!10}
                                     & schubfach          & 15.0  & 340   & 4.1   & 8.4   & 310   & 7.4 \\
    \rowcolor{white}
    gaia (f64)                       & dragonbox          & 22.0  & 380   & 3.2   & 14.0  & 250   & 4.1 \\
    \rowcolor{white}
                                     & ryu                & 26.0  & 540   & 3.7   & 16.0  & 320   & 4.8 \\
    \rowcolor{white}
                                     & schubfach          & 26.0  & 460   & 3.3   & 16.0  & 310   & 4.8 \\
    \rowcolor{black!10}
    marine\_ik (f32)                 & dragonbox          & 9.3   & 210   & 4.2   & 5.7   & 180   & 7.1 \\
    \rowcolor{black!10}
                                     & ryu                & 14.0  & 370   & 4.6   & 8.2   & 270   & 7.2 \\
    \rowcolor{black!10}
                                     & schubfach          & 10.0  & 260   & 4.6   & 6.3   & 230   & 8.2 \\
    \rowcolor{white}
    mobilenetv3\_large (f32)         & dragonbox          & 18.0  & 220   & 2.2   & 14.0  & 180   & 3.1 \\
    \rowcolor{white}
                                     & ryu                & 19.0  & 350   & 3.3   & 14.0  & 250   & 4.3 \\
    \rowcolor{white}
                                     & schubfach          & 17.0  & 270   & 2.8   & 13.0  & 240   & 4.4 \\
    \rowcolor{black!10}
    noaa\_gfs\_1p00 (f32)            & dragonbox          & 12.0  & 190   & 2.9   & 9.1   & 160   & 4.2 \\
    \rowcolor{black!10}
                                     & ryu                & 15.0  & 310   & 3.8   & 11.0  & 220   & 5.0 \\
    \rowcolor{black!10}
                                     & schubfach          & 13.0  & 240   & 3.3   & 9.0   & 210   & 5.5 \\
    \rowcolor{white}
    noaa\_global\_hourly\_2023 (f32) & dragonbox          & 10.0  & 210   & 3.7   & 7.7   & 170   & 5.1 \\
    \rowcolor{white}
                                     & ryu                & 17.0  & 400   & 4.3   & 13.0  & 310   & 5.6 \\
    \rowcolor{white}
                                     & schubfach          & 8.9   & 220   & 4.4   & 6.8   & 190   & 6.6 \\
    \bottomrule
  \end{tabular}\restartrowcolors
\end{table*}

\section{Conclusion}%
\label{sec:conclusion}

Although the original Dragon4 algorithm required thousands of instructions to
convert a single IEEE floating-point number to a decimal string, modern methods
accomplish the same task in just a few hundred instructions. This tenfold
improvement over three decades represents an average software efficiency gain of
about \SI{8}{\percent} per year---a striking reminder that performance gains
in software, like those in hardware, can accumulate to significant effect over
time~\cite{leiserson2020there}.

Yet, despite these advances, our results show that widely used libraries (such
as \texttt{fmt} and \texttt{libc++}) still require more instructions than
state-of-the-art algorithms like Dragonbox. This suggests that these libraries
may be prioritizing other trade-offs (e.g., code size or portability), or that
further performance optimizations remain unexplored. Moreover, while each
algorithm we tested produces a valid string representation, none consistently
generates the shortest possible output. We identify two key directions for future research:
\begin{itemize}
  \item \emph{High performance string generation.} Historically, converting
    decimal significands and exponents into ASCII strings was a minor portion of
    the overall float-to-string process. However, as core conversion algorithms
    have become dramatically faster, this final string generation step can now
    consume a significant share of runtime---for example, only 2\% of cycles in
    Dragon4 on the unit dataset, but up to 34\% in \texttt{std::to\_chars}.
    Future work should explore advanced optimizations for this stage, ideally
    decoupled from earlier steps, to further accelerate end-to-end conversion.
    Related to this, several modern algorithms (e.g., Dragonbox, Schubfach, and
    Ryu) efficiently compute the shortest decimal significand but leave the
    final construction of the shortest string to user-defined routines.
    Developing unified backends that consistently generate the shortest possible
    decimal string across algorithms represents a promising direction for future work.
  \item \emph{Exploiting modern CPU features.} Our results show that current
    algorithms do not fully utilize advanced instructions such as FMA and SIMD,
    and that enabling SIMD extensions (e.g., AVX-512) yields, at best, marginal
    gains for single-value conversions. Future work should investigate whether
    these capabilities can be leveraged by reformulating the problem---for
    example, by designing algorithms that convert multiple floats to strings in
    parallel (batch conversion). Such an approach could better exploit
    vectorization and other modern CPU features, potentially unlocking
    significant performance gains for applications that require bulk formatting
    of floating-point data.
\end{itemize}
Also, as future work, we might investigate the generation of the significand and
exponent independently from the generation of the string.

\section*{Author Contributions}

Jaël Champagne Gareau: conceptualization; investigation; software; experimentation; writing-review and editing.
Daniel Lemire: conceptualization; software; validation; experimentation; data analysis; writing-original draft; writing-review and editing.

\section*{Funding Information}

This work was supported by the Natural Sciences and Engineering Research Council of Canada, Grant Number: RGPIN-2017-03910.
The first author is supported by a postdoctoral grant from Fonds de recherche du Québec, \url{https://doi.org/10.69777/361128}.

\section*{Data Availability Statement}

All our data and software is freely available online. The C++ benchmarking software is available online
at \url{https://github.com/fastfloat/float_serialization_benchmark}. All of our test datasets are at \url{https://github.com/fastfloat/float-data}.
We collected performance data and we make it available on the paper's webpage at \url{https://www.jaelgareau.com/en/publication/gareau_lemire-spe25/}.

\bibliography{ref}

\end{document}


\input{AMD_Ryzen9_9900X/Ryzen9900x_g++_all_none.tex}
\input{AMD_Ryzen9_9900X/Ryzen9900x_clang++_all_none.tex}
\input{AMD_Ryzen9_9900X/Ryzen9900x_g++_all_s.tex}
\input{AMD_Ryzen9_9900X/Ryzen9900x_clang++_all_s.tex}

\input{c6g.medium/Neoverse-N1_g++_all_none.tex}
\input{c6g.medium/Neoverse-N1_clang++_all_none.tex}
\input{c6g.medium/Neoverse-N1_g++_all_s.tex}
\input{c6g.medium/Neoverse-N1_clang++_all_s.tex}

\input{c7g.medium/Neoverse-V1_g++_all_none.tex}
\input{c7g.medium/Neoverse-V1_clang++_all_none.tex}
\input{c7g.medium/Neoverse-V1_g++_all_s.tex}
\input{c7g.medium/Neoverse-V1_clang++_all_s.tex}

\input{c8g.medium/Neoverse-V2_g++_all_none.tex}
\input{c8g.medium/Neoverse-V2_clang++_all_none.tex}
\input{c8g.medium/Neoverse-V2_g++_all_s.tex}
\input{c8g.medium/Neoverse-V2_clang++_all_s.tex}

\input{c7a.metal-48xl/AMD_EPYC_9R14_96-Core_Processor_g++_all_none.tex}
\input{c7a.metal-48xl/AMD_EPYC_9R14_96-Core_Processor_clang++_all_none.tex}
\input{c7a.metal-48xl/AMD_EPYC_9R14_96-Core_Processor_g++_all_s.tex}
\input{c7a.metal-48xl/AMD_EPYC_9R14_96-Core_Processor_clang++_all_s.tex}

\input{c6i.metal/Intel(R)_Xeon(R)_Platinum_8375C_CPU__2.90GHz_g++_all_none.tex}
\input{c6i.metal/Intel(R)_Xeon(R)_Platinum_8375C_CPU__2.90GHz_clang++_all_none.tex}
\input{c6i.metal/Intel(R)_Xeon(R)_Platinum_8375C_CPU__2.90GHz_g++_all_s.tex}
\input{c6i.metal/Intel(R)_Xeon(R)_Platinum_8375C_CPU__2.90GHz_clang++_all_s.tex}

\input{c7i.metal-24xl/Intel(R)_Xeon(R)_Platinum_8488C_g++_all_none.tex}
\input{c7i.metal-24xl/Intel(R)_Xeon(R)_Platinum_8488C_clang++_all_none.tex}
\input{c7i.metal-24xl/Intel(R)_Xeon(R)_Platinum_8488C_g++_all_s.tex}
\input{c7i.metal-24xl/Intel(R)_Xeon(R)_Platinum_8488C_clang++_all_s.tex}

\input{c6a.metal/AMD_EPYC_7R13_48-Core_Processor_g++_all_none.tex}
\input{c6a.metal/AMD_EPYC_7R13_48-Core_Processor_clang++_all_none.tex}
\input{c6a.metal/AMD_EPYC_7R13_48-Core_Processor_g++_all_s.tex}
\input{c6a.metal/AMD_EPYC_7R13_48-Core_Processor_clang++_all_s.tex}

\input{c5n.metal/Intel(R)_Xeon(R)_Platinum_8124M_CPU__3.00GHz_g++_all_none.tex}
\input{c5n.metal/Intel(R)_Xeon(R)_Platinum_8124M_CPU__3.00GHz_clang++_all_none.tex}
\input{c5n.metal/Intel(R)_Xeon(R)_Platinum_8124M_CPU__3.00GHz_g++_all_s.tex}
\input{c5n.metal/Intel(R)_Xeon(R)_Platinum_8124M_CPU__3.00GHz_clang++_all_s.tex}

\input{c5a.24xlarge/AMD_EPYC_7R32_g++_all_none.tex}
\input{c5a.24xlarge/AMD_EPYC_7R32_clang++_all_none.tex}
\input{c5a.24xlarge/AMD_EPYC_7R32_g++_all_s.tex}
\input{c5a.24xlarge/AMD_EPYC_7R32_clang++_all_s.tex}

\input{apple_m4/Apple_M4_Max_clang++_all_none.tex}
\input{apple_m4/Apple_M4_Max_clang++_all_s.tex}

%% file: results/AMD_Ryzen9_9900X/Ryzen9900x_g++_all_none.tex
\begin{table}
  \centering
  \caption{Ryzen 9900X results (g++, 64-bit floats)}%
  \label{tab:ryzen9900xresults}
  \begin{tabular}{lccccccccc}
    \toprule
    \multirow{1}{*}{Name}   & \multicolumn{3}{c|}{mesh}  & \multicolumn{3}{c|}{canada} & \multicolumn{3}{c}{unit} \\
                            & {ns/f} & {ins/f} & {ins/c} & {ns/f} & {ins/f} & {ins/c}  & {ns/f} & {ins/f} & {ins/c} \\ \midrule
    dragon4 & 82 & 2300 & 5.0  & 170 & 4700 & 5.2  & 190 & 5000 & 4.9  \\
    fmt\_format & 30 & 570 & 3.5  & 40 & 840 & 3.8  & 35 & 560 & 2.9  \\
    grisu3 & 12 & 290 & 4.5  & 29 & 630 & 4.0  & 26 & 510 & 3.6  \\
    grisu\_exact & 18 & 370 & 3.7  & 24 & 520 & 3.9  & 21 & 370 & 3.1  \\
    schubfach & 9.9 & 250 & 4.5  & 24 & 490 & 3.7  & 19 & 320 & 3.0  \\
    dragonbox & 11 & 260 & 4.3  & 18 & 410 & 4.1  & 15 & 240 & 3.0  \\
    ryu & 14 & 320 & 4.3  & 24 & 580 & 4.3  & 20 & 400 & 3.5  \\
    double\_conversion & 27 & 610 & 4.0  & 45 & 1000 & 4.0  & 39 & 810 & 3.7  \\
    swiftDtoa & 23 & 490 & 3.8  & 28 & 590 & 3.8  & 27 & 440 & 3.0  \\
    std::to\_chars & 18 & 490 & 4.8  & 30 & 780 & 4.8  & 25 & 600 & 4.3  \\
    \bottomrule
  \end{tabular}\restartrowcolors
\end{table}

%% file: results/AMD_Ryzen9_9900X/Ryzen9900x_clang++_all_none.tex
\begin{table}
  \centering
  \caption{Ryzen 9900X results (clang++, 64-bit floats)}%
  \label{tab:ryzen9900xresults}
  \begin{tabular}{lccccccccc}
    \toprule
    \multirow{1}{*}{Name}   & \multicolumn{3}{c|}{mesh}  & \multicolumn{3}{c|}{canada} & \multicolumn{3}{c}{unit} \\
                            & {ns/f} & {ins/f} & {ins/c} & {ns/f} & {ins/f} & {ins/c}  & {ns/f} & {ins/f} & {ins/c} \\ \midrule
    dragon4 & 76 & 2200 & 5.3  & 160 & 4500 & 5.1  & 170 & 4700 & 5.2  \\
    fmt\_format & 27 & 580 & 3.9  & 35 & 840 & 4.4  & 30 & 530 & 3.2  \\
    grisu3 & 12 & 290 & 4.4  & 28 & 600 & 3.8  & 25 & 470 & 3.3  \\
    grisu\_exact & 16 & 340 & 3.9  & 22 & 480 & 3.8  & 19 & 320 & 3.0  \\
    schubfach & 9.9 & 240 & 4.4  & 23 & 460 & 3.7  & 18 & 290 & 2.9  \\
    dragonbox & 12 & 250 & 3.7  & 18 & 390 & 4.1  & 16 & 230 & 2.5  \\
    ryu & 13 & 320 & 4.3  & 22 & 530 & 4.4  & 18 & 340 & 3.5  \\
    double\_conversion & 31 & 740 & 4.4  & 50 & 1200 & 4.2  & 44 & 980 & 4.0  \\
    swiftDtoa & 20 & 440 & 4.0  & 24 & 530 & 4.0  & 23 & 380 & 3.0  \\
    std::to\_chars & 19 & 500 & 4.9  & 29 & 790 & 4.9  & 25 & 600 & 4.3  \\
    \bottomrule
  \end{tabular}\restartrowcolors
\end{table}

%% file: results/AMD_Ryzen9_9900X/Ryzen9900x_g++_all_s.tex
\begin{table}
  \centering
  \caption{Ryzen 9900X results (g++, 32-bit floats)}%
  \label{tab:ryzen9900xresults}
  \begin{tabular}{lccccccccc}
    \toprule
    \multirow{1}{*}{Name}   & \multicolumn{3}{c|}{mesh}  & \multicolumn{3}{c|}{canada} & \multicolumn{3}{c}{unit} \\
                            & {ns/f} & {ins/f} & {ins/c} & {ns/f} & {ins/f} & {ins/c}  & {ns/f} & {ins/f} & {ins/c} \\ \midrule
    dragon4 & 62 & 1700 & 5.0  & 84 & 2400 & 5.3  & 99 & 2400 & 4.4  \\
    fmt\_format & 23 & 470 & 3.6  & 32 & 710 & 4.0  & 25 & 410 & 3.0  \\
    grisu\_exact & 16 & 310 & 3.5  & 24 & 430 & 3.4  & 19 & 270 & 2.5  \\
    schubfach & 9.8 & 210 & 3.9  & 18 & 410 & 4.0  & 14 & 230 & 2.9  \\
    dragonbox & 10 & 200 & 3.6  & 18 & 340 & 3.5  & 13 & 170 & 2.4  \\
    ryu & 16 & 390 & 4.3  & 20 & 480 & 4.3  & 17 & 310 & 3.2  \\
    double\_conversion & 27 & 610 & 4.1  & 38 & 860 & 4.1  & 34 & 660 & 3.6  \\
    swiftDtoa & 15 & 300 & 3.6  & 20 & 460 & 4.1  & 21 & 300 & 2.6  \\
    std::to\_chars & 21 & 540 & 4.6  & 26 & 630 & 4.4  & 23 & 450 & 3.6  \\
    \bottomrule
  \end{tabular}\restartrowcolors
\end{table}

%% file: results/AMD_Ryzen9_9900X/Ryzen9900x_clang++_all_s.tex
\begin{table}
  \centering
  \caption{Ryzen 9900X results (clang++, 32-bit floats)}%
  \label{tab:ryzen9900xresults}
  \begin{tabular}{lccccccccc}
    \toprule
    \multirow{1}{*}{Name}   & \multicolumn{3}{c|}{mesh}  & \multicolumn{3}{c|}{canada} & \multicolumn{3}{c}{unit} \\
                            & {ns/f} & {ins/f} & {ins/c} & {ns/f} & {ins/f} & {ins/c}  & {ns/f} & {ins/f} & {ins/c} \\ \midrule
    dragon4 & 54 & 1600 & 5.6  & 78 & 2400 & 5.6  & 81 & 2300 & 5.2  \\
    fmt\_format & 23 & 490 & 3.9  & 33 & 730 & 4.1  & 24 & 420 & 3.2  \\
    grisu\_exact & 15 & 290 & 3.6  & 21 & 430 & 3.6  & 19 & 260 & 2.5  \\
    schubfach & 9.8 & 220 & 4.0  & 18 & 400 & 4.0  & 15 & 220 & 2.7  \\
    dragonbox & 10 & 210 & 3.6  & 17 & 340 & 3.8  & 14 & 180 & 2.3  \\
    ryu & 16 & 380 & 4.2  & 20 & 450 & 4.2  & 17 & 280 & 3.0  \\
    double\_conversion & 31 & 740 & 4.3  & 43 & 1000 & 4.3  & 38 & 820 & 3.9  \\
    swiftDtoa & 15 & 320 & 3.8  & 20 & 470 & 4.2  & 21 & 320 & 2.7  \\
    std::to\_chars & 21 & 550 & 4.6  & 25 & 630 & 4.6  & 23 & 460 & 3.6  \\
    \bottomrule
  \end{tabular}\restartrowcolors
\end{table}

%% file: results/c6g.medium/Neoverse-N1_g++_all_none.tex
\begin{table}
  \centering
  \caption{Neoverse-N1 results (g++, 64-bit floats)}%
  \label{tab:neoversen1results}
  \begin{tabular}{lccccccccc}
    \toprule
    \multirow{1}{*}{Name}   & \multicolumn{3}{c|}{mesh}  & \multicolumn{3}{c|}{canada} & \multicolumn{3}{c}{unit} \\
                            & {ns/f} & {ins/f} & {ins/c} & {ns/f} & {ins/f} & {ins/c}  & {ns/f} & {ins/f} & {ins/c} \\ \midrule
    dragon4 & 220 & 1700 & 3.2  & 460 & 3600 & 3.1  & 460 & 3700 & 3.2  \\
    fmt\_format & 88 & 610 & 2.8  & 170 & 1200 & 2.9  & 94 & 590 & 2.5  \\
    grisu3 & 40 & 310 & 3.1  & 99 & 700 & 2.8  & 78 & 500 & 2.6  \\
    grisu\_exact & 67 & 410 & 2.4  & 93 & 610 & 2.6  & 66 & 380 & 2.3  \\
    schubfach & 35 & 250 & 2.9  & 85 & 540 & 2.6  & 56 & 300 & 2.1  \\
    dragonbox & 39 & 260 & 2.7  & 67 & 470 & 2.8  & 43 & 230 & 2.2  \\
    ryu & 49 & 330 & 2.7  & 87 & 600 & 2.8  & 57 & 360 & 2.5  \\
    double\_conversion & 88 & 610 & 2.8  & 160 & 1000 & 2.7  & 120 & 790 & 2.7  \\
    swiftDtoa & 66 & 410 & 2.5  & 91 & 570 & 2.5  & 66 & 360 & 2.2  \\
    std::to\_chars & 71 & 490 & 2.8  & 120 & 810 & 2.7  & 87 & 570 & 2.6  \\
    \bottomrule
  \end{tabular}\restartrowcolors
\end{table}

%% file: results/c6g.medium/Neoverse-N1_clang++_all_none.tex
\begin{table}
  \centering
  \caption{Neoverse-N1 results (clang++, 64-bit floats)}%
  \label{tab:neoversen1results}
  \begin{tabular}{lccccccccc}
    \toprule
    \multirow{1}{*}{Name}   & \multicolumn{3}{c|}{mesh}  & \multicolumn{3}{c|}{canada} & \multicolumn{3}{c}{unit} \\
                            & {ns/f} & {ins/f} & {ins/c} & {ns/f} & {ins/f} & {ins/c}  & {ns/f} & {ins/f} & {ins/c} \\ \midrule
    dragon4 & 200 & 1600 & 3.1  & 430 & 3300 & 3.1  & 440 & 3400 & 3.1  \\
    fmt\_format & 85 & 530 & 2.5  & 140 & 820 & 2.4  & 89 & 490 & 2.2  \\
    grisu3 & 40 & 290 & 2.9  & 100 & 650 & 2.6  & 79 & 470 & 2.4  \\
    grisu\_exact & 58 & 340 & 2.4  & 83 & 550 & 2.7  & 57 & 330 & 2.3  \\
    schubfach & 33 & 240 & 2.9  & 85 & 520 & 2.5  & 57 & 290 & 2.0  \\
    dragonbox & 40 & 250 & 2.6  & 66 & 450 & 2.8  & 42 & 230 & 2.2  \\
    ryu & 50 & 290 & 2.3  & 85 & 540 & 2.6  & 57 & 310 & 2.2  \\
    double\_conversion & 90 & 650 & 2.9  & 160 & 1100 & 2.8  & 120 & 870 & 2.8  \\
    swiftDtoa & 65 & 400 & 2.5  & 89 & 560 & 2.6  & 66 & 360 & 2.2  \\
    std::to\_chars & 69 & 480 & 2.8  & 120 & 810 & 2.6  & 87 & 570 & 2.6  \\
    \bottomrule
  \end{tabular}\restartrowcolors
\end{table}

%% file: results/c6g.medium/Neoverse-N1_g++_all_s.tex
\begin{table}
  \centering
  \caption{Neoverse-N1 results (g++, 32-bit floats)}%
  \label{tab:neoversen1results}
  \begin{tabular}{lccccccccc}
    \toprule
    \multirow{1}{*}{Name}   & \multicolumn{3}{c|}{mesh}  & \multicolumn{3}{c|}{canada} & \multicolumn{3}{c}{unit} \\
                            & {ns/f} & {ins/f} & {ins/c} & {ns/f} & {ins/f} & {ins/c}  & {ns/f} & {ins/f} & {ins/c} \\ \midrule
    dragon4 & 170 & 1300 & 3.2  & 260 & 1900 & 3.0  & 240 & 1800 & 3.1  \\
    fmt\_format & 72 & 530 & 3.0  & 120 & 870 & 2.9  & 70 & 460 & 2.6  \\
    grisu\_exact & 43 & 320 & 3.0  & 74 & 510 & 2.8  & 46 & 280 & 2.4  \\
    schubfach & 31 & 220 & 3.0  & 68 & 470 & 2.8  & 42 & 230 & 2.2  \\
    dragonbox & 30 & 210 & 2.8  & 57 & 410 & 2.9  & 31 & 170 & 2.2  \\
    ryu & 48 & 350 & 3.0  & 70 & 500 & 2.9  & 44 & 270 & 2.5  \\
    double\_conversion & 91 & 610 & 2.7  & 140 & 900 & 2.6  & 100 & 650 & 2.6  \\
    swiftDtoa & 45 & 290 & 2.6  & 73 & 500 & 2.7  & 50 & 290 & 2.3  \\
    std::to\_chars & 66 & 500 & 3.1  & 87 & 650 & 3.0  & 58 & 410 & 2.8  \\
    \bottomrule
  \end{tabular}\restartrowcolors
\end{table}

%% file: results/c6g.medium/Neoverse-N1_clang++_all_s.tex
\begin{table}
  \centering
  \caption{Neoverse-N1 results (clang++, 32-bit floats)}%
  \label{tab:neoversen1results}
  \begin{tabular}{lccccccccc}
    \toprule
    \multirow{1}{*}{Name}   & \multicolumn{3}{c|}{mesh}  & \multicolumn{3}{c|}{canada} & \multicolumn{3}{c}{unit} \\
                            & {ns/f} & {ins/f} & {ins/c} & {ns/f} & {ins/f} & {ins/c}  & {ns/f} & {ins/f} & {ins/c} \\ \midrule
    dragon4 & 160 & 1200 & 3.1  & 230 & 1800 & 3.1  & 230 & 1700 & 3.0  \\
    fmt\_format & 59 & 450 & 3.1  & 95 & 720 & 3.1  & 55 & 390 & 2.8  \\
    grisu\_exact & 37 & 270 & 3.0  & 63 & 460 & 3.0  & 42 & 240 & 2.3  \\
    schubfach & 30 & 210 & 2.9  & 64 & 460 & 2.9  & 41 & 220 & 2.2  \\
    dragonbox & 30 & 210 & 2.8  & 55 & 400 & 2.9  & 31 & 170 & 2.2  \\
    ryu & 42 & 330 & 3.1  & 60 & 470 & 3.1  & 39 & 240 & 2.5  \\
    double\_conversion & 93 & 650 & 2.8  & 140 & 960 & 2.7  & 110 & 710 & 2.7  \\
    swiftDtoa & 43 & 300 & 2.8  & 67 & 500 & 3.0  & 50 & 300 & 2.4  \\
    std::to\_chars & 65 & 500 & 3.1  & 83 & 640 & 3.1  & 57 & 410 & 2.9  \\
    \bottomrule
  \end{tabular}\restartrowcolors
\end{table}

%% file: results/c7g.medium/Neoverse-V1_g++_all_none.tex
\begin{table}
  \centering
  \caption{Neoverse-V1 results (g++, 64-bit floats)}%
  \label{tab:neoversev1results}
  \begin{tabular}{lccccccccc}
    \toprule
    \multirow{1}{*}{Name}   & \multicolumn{3}{c|}{mesh}  & \multicolumn{3}{c|}{canada} & \multicolumn{3}{c}{unit} \\
                            & {ns/f} & {ins/f} & {ins/c} & {ns/f} & {ins/f} & {ins/c}  & {ns/f} & {ins/f} & {ins/c} \\ \midrule
    dragon4 & 150 & 1700 & 4.5  & 300 & 3500 & 4.5  & 320 & 3600 & 4.4  \\
    fmt\_format & 59 & 610 & 4.0  & 110 & 1200 & 4.2  & 63 & 590 & 3.6  \\
    grisu3 & 27 & 310 & 4.4  & 65 & 690 & 4.1  & 55 & 500 & 3.5  \\
    grisu\_exact & 38 & 400 & 4.1  & 56 & 610 & 4.2  & 42 & 380 & 3.4  \\
    schubfach & 20 & 250 & 4.7  & 49 & 540 & 4.3  & 34 & 300 & 3.3  \\
    dragonbox & 24 & 250 & 4.1  & 41 & 460 & 4.3  & 28 & 230 & 3.2  \\
    ryu & 30 & 320 & 4.2  & 51 & 600 & 4.5  & 36 & 360 & 3.8  \\
    double\_conversion & 59 & 600 & 3.9  & 100 & 1000 & 3.8  & 81 & 780 & 3.7  \\
    swiftDtoa & 43 & 410 & 3.7  & 52 & 560 & 4.2  & 43 & 360 & 3.2  \\
    std::to\_chars & 42 & 480 & 4.5  & 72 & 810 & 4.4  & 53 & 570 & 4.1  \\
    \bottomrule
  \end{tabular}\restartrowcolors
\end{table}

%% file: results/c7g.medium/Neoverse-V1_clang++_all_none.tex
\begin{table}
  \centering
  \caption{Neoverse-V1 results (clang++, 64-bit floats)}%
  \label{tab:neoversev1results}
  \begin{tabular}{lccccccccc}
    \toprule
    \multirow{1}{*}{Name}   & \multicolumn{3}{c|}{mesh}  & \multicolumn{3}{c|}{canada} & \multicolumn{3}{c}{unit} \\
                            & {ns/f} & {ins/f} & {ins/c} & {ns/f} & {ins/f} & {ins/c}  & {ns/f} & {ins/f} & {ins/c} \\ \midrule
    dragon4 & 140 & 1600 & 4.4  & 290 & 3300 & 4.4  & 310 & 3400 & 4.3  \\
    fmt\_format & 59 & 530 & 3.5  & 85 & 820 & 3.7  & 56 & 490 & 3.4  \\
    grisu3 & 30 & 290 & 3.8  & 67 & 650 & 3.7  & 56 & 470 & 3.2  \\
    grisu\_exact & 35 & 340 & 3.7  & 54 & 550 & 3.9  & 41 & 330 & 3.1  \\
    schubfach & 22 & 230 & 4.1  & 50 & 520 & 4.0  & 36 & 290 & 3.1  \\
    dragonbox & 26 & 250 & 3.7  & 41 & 450 & 4.2  & 27 & 230 & 3.2  \\
    ryu & 29 & 290 & 3.7  & 51 & 540 & 4.1  & 35 & 310 & 3.4  \\
    double\_conversion & 68 & 640 & 3.6  & 120 & 1100 & 3.6  & 93 & 870 & 3.6  \\
    swiftDtoa & 43 & 400 & 3.6  & 53 & 560 & 4.1  & 43 & 360 & 3.2  \\
    std::to\_chars & 42 & 480 & 4.4  & 69 & 800 & 4.5  & 53 & 570 & 4.1  \\
    \bottomrule
  \end{tabular}\restartrowcolors
\end{table}

%% file: results/c7g.medium/Neoverse-V1_g++_all_s.tex
\begin{table}
  \centering
  \caption{Neoverse-V1 results (g++, 32-bit floats)}%
  \label{tab:neoversev1results}
  \begin{tabular}{lccccccccc}
    \toprule
    \multirow{1}{*}{Name}   & \multicolumn{3}{c|}{mesh}  & \multicolumn{3}{c|}{canada} & \multicolumn{3}{c}{unit} \\
                            & {ns/f} & {ins/f} & {ins/c} & {ns/f} & {ins/f} & {ins/c}  & {ns/f} & {ins/f} & {ins/c} \\ \midrule
    dragon4 & 120 & 1300 & 4.4  & 170 & 1900 & 4.3  & 170 & 1800 & 4.0  \\
    fmt\_format & 54 & 520 & 3.7  & 89 & 870 & 3.8  & 51 & 460 & 3.5  \\
    grisu\_exact & 29 & 310 & 4.2  & 46 & 500 & 4.2  & 34 & 280 & 3.2  \\
    schubfach & 19 & 220 & 4.5  & 40 & 470 & 4.6  & 26 & 230 & 3.5  \\
    dragonbox & 20 & 200 & 3.9  & 36 & 410 & 4.4  & 22 & 170 & 3.0  \\
    ryu & 33 & 350 & 4.2  & 43 & 500 & 4.5  & 30 & 270 & 3.4  \\
    double\_conversion & 60 & 610 & 3.9  & 88 & 890 & 3.9  & 69 & 650 & 3.6  \\
    swiftDtoa & 29 & 290 & 3.8  & 44 & 490 & 4.3  & 36 & 290 & 3.1  \\
    std::to\_chars & 44 & 500 & 4.4  & 54 & 640 & 4.6  & 42 & 410 & 3.8  \\
    \bottomrule
  \end{tabular}\restartrowcolors
\end{table}

%% file: results/c7g.medium/Neoverse-V1_clang++_all_s.tex
\begin{table}
  \centering
  \caption{Neoverse-V1 results (clang++, 32-bit floats)}%
  \label{tab:neoversev1results}
  \begin{tabular}{lccccccccc}
    \toprule
    \multirow{1}{*}{Name}   & \multicolumn{3}{c|}{mesh}  & \multicolumn{3}{c|}{canada} & \multicolumn{3}{c}{unit} \\
                            & {ns/f} & {ins/f} & {ins/c} & {ns/f} & {ins/f} & {ins/c}  & {ns/f} & {ins/f} & {ins/c} \\ \midrule
    dragon4 & 110 & 1200 & 4.3  & 160 & 1800 & 4.2  & 160 & 1700 & 4.0  \\
    fmt\_format & 46 & 450 & 3.8  & 66 & 720 & 4.3  & 40 & 390 & 3.8  \\
    grisu\_exact & 28 & 270 & 3.7  & 44 & 460 & 4.0  & 33 & 240 & 2.8  \\
    schubfach & 20 & 210 & 4.0  & 41 & 450 & 4.2  & 29 & 220 & 3.0  \\
    dragonbox & 21 & 200 & 3.7  & 37 & 400 & 4.1  & 23 & 170 & 3.0  \\
    ryu & 33 & 320 & 3.8  & 41 & 460 & 4.3  & 29 & 240 & 3.2  \\
    double\_conversion & 69 & 650 & 3.6  & 98 & 950 & 3.7  & 79 & 710 & 3.5  \\
    swiftDtoa & 31 & 290 & 3.6  & 42 & 490 & 4.5  & 35 & 300 & 3.2  \\
    std::to\_chars & 44 & 500 & 4.4  & 53 & 640 & 4.6  & 42 & 410 & 3.8  \\
    \bottomrule
  \end{tabular}\restartrowcolors
\end{table}

%% file: results/c8g.medium/Neoverse-V2_g++_all_none.tex
\begin{table}
  \centering
  \caption{Neoverse-V2 results (g++, 64-bit floats)}%
  \label{tab:neoversev2results}
  \begin{tabular}{lccccccccc}
    \toprule
    \multirow{1}{*}{Name}   & \multicolumn{3}{c|}{mesh}  & \multicolumn{3}{c|}{canada} & \multicolumn{3}{c}{unit} \\
                            & {ns/f} & {ins/f} & {ins/c} & {ns/f} & {ins/f} & {ins/c}  & {ns/f} & {ins/f} & {ins/c} \\ \midrule
    dragon4 & 110 & 1700 & 5.5  & 230 & 3500 & 5.6  & 240 & 3700 & 5.4  \\
    fmt\_format & 49 & 610 & 4.4  & 84 & 1200 & 5.1  & 52 & 590 & 4.0  \\
    grisu3 & 21 & 310 & 5.2  & 51 & 690 & 4.8  & 45 & 500 & 4.0  \\
    grisu\_exact & 28 & 400 & 5.1  & 45 & 610 & 4.8  & 33 & 380 & 4.1  \\
    schubfach & 15 & 250 & 5.8  & 38 & 540 & 5.0  & 28 & 300 & 3.9  \\
    dragonbox & 17 & 250 & 5.2  & 33 & 460 & 5.0  & 23 & 230 & 3.6  \\
    ryu & 21 & 320 & 5.4  & 42 & 600 & 5.2  & 29 & 360 & 4.4  \\
    double\_conversion & 46 & 600 & 4.6  & 84 & 1000 & 4.4  & 68 & 780 & 4.1  \\
    swiftDtoa & 34 & 410 & 4.3  & 42 & 560 & 4.8  & 35 & 360 & 3.7  \\
    std::to\_chars & 32 & 480 & 5.4  & 55 & 810 & 5.2  & 41 & 570 & 5.0  \\
    \bottomrule
  \end{tabular}\restartrowcolors
\end{table}

%% file: results/c8g.medium/Neoverse-V2_clang++_all_none.tex
\begin{table}
  \centering
  \caption{Neoverse-V2 results (clang++, 64-bit floats)}%
  \label{tab:neoversev2results}
  \begin{tabular}{lccccccccc}
    \toprule
    \multirow{1}{*}{Name}   & \multicolumn{3}{c|}{mesh}  & \multicolumn{3}{c|}{canada} & \multicolumn{3}{c}{unit} \\
                            & {ns/f} & {ins/f} & {ins/c} & {ns/f} & {ins/f} & {ins/c}  & {ns/f} & {ins/f} & {ins/c} \\ \midrule
    dragon4 & 110 & 1600 & 5.4  & 220 & 3300 & 5.3  & 230 & 3400 & 5.3  \\
    fmt\_format & 44 & 530 & 4.3  & 68 & 830 & 4.4  & 48 & 490 & 3.7  \\
    grisu3 & 21 & 290 & 4.9  & 52 & 650 & 4.4  & 45 & 470 & 3.8  \\
    grisu\_exact & 27 & 340 & 4.6  & 44 & 550 & 4.4  & 32 & 330 & 3.7  \\
    schubfach & 16 & 230 & 5.3  & 41 & 520 & 4.6  & 28 & 290 & 3.6  \\
    dragonbox & 19 & 250 & 4.8  & 33 & 450 & 5.0  & 23 & 230 & 3.6  \\
    ryu & 21 & 290 & 4.8  & 40 & 540 & 4.8  & 28 & 310 & 4.0  \\
    double\_conversion & 55 & 640 & 4.2  & 93 & 1100 & 4.2  & 77 & 870 & 4.0  \\
    swiftDtoa & 31 & 400 & 4.5  & 42 & 560 & 4.8  & 34 & 360 & 3.8  \\
    std::to\_chars & 31 & 480 & 5.5  & 54 & 800 & 5.4  & 40 & 570 & 5.0  \\
    \bottomrule
  \end{tabular}\restartrowcolors
\end{table}

%% file: results/c8g.medium/Neoverse-V2_g++_all_s.tex
\begin{table}
  \centering
  \caption{Neoverse-V2 results (g++, 32-bit floats)}%
  \label{tab:neoversev2results}
  \begin{tabular}{lccccccccc}
    \toprule
    \multirow{1}{*}{Name}   & \multicolumn{3}{c|}{mesh}  & \multicolumn{3}{c|}{canada} & \multicolumn{3}{c}{unit} \\
                            & {ns/f} & {ins/f} & {ins/c} & {ns/f} & {ins/f} & {ins/c}  & {ns/f} & {ins/f} & {ins/c} \\ \midrule
    dragon4 & 87 & 1300 & 5.5  & 130 & 1900 & 5.3  & 130 & 1800 & 5.1  \\
    fmt\_format & 47 & 520 & 4.0  & 70 & 870 & 4.5  & 44 & 460 & 3.7  \\
    grisu\_exact & 22 & 310 & 5.2  & 37 & 500 & 4.8  & 27 & 280 & 3.6  \\
    schubfach & 14 & 220 & 5.5  & 32 & 470 & 5.2  & 21 & 230 & 3.9  \\
    dragonbox & 15 & 200 & 5.0  & 30 & 410 & 4.9  & 19 & 170 & 3.2  \\
    ryu & 25 & 350 & 5.0  & 34 & 500 & 5.2  & 25 & 270 & 3.9  \\
    double\_conversion & 47 & 610 & 4.6  & 72 & 890 & 4.4  & 55 & 650 & 4.2  \\
    swiftDtoa & 22 & 290 & 4.7  & 36 & 490 & 4.9  & 30 & 290 & 3.5  \\
    std::to\_chars & 34 & 500 & 5.2  & 46 & 640 & 5.0  & 34 & 410 & 4.3  \\
    \bottomrule
  \end{tabular}\restartrowcolors
\end{table}

%% file: results/c8g.medium/Neoverse-V2_clang++_all_s.tex
\begin{table}
  \centering
  \caption{Neoverse-V2 results (clang++, 32-bit floats)}%
  \label{tab:neoversev2results}
  \begin{tabular}{lccccccccc}
    \toprule
    \multirow{1}{*}{Name}   & \multicolumn{3}{c|}{mesh}  & \multicolumn{3}{c|}{canada} & \multicolumn{3}{c}{unit} \\
                            & {ns/f} & {ins/f} & {ins/c} & {ns/f} & {ins/f} & {ins/c}  & {ns/f} & {ins/f} & {ins/c} \\ \midrule
    dragon4 & 81 & 1200 & 5.4  & 130 & 1800 & 5.1  & 120 & 1700 & 5.0  \\
    fmt\_format & 33 & 450 & 4.9  & 54 & 730 & 4.8  & 33 & 390 & 4.2  \\
    grisu\_exact & 21 & 270 & 4.5  & 37 & 460 & 4.5  & 28 & 240 & 3.1  \\
    schubfach & 15 & 210 & 4.9  & 32 & 450 & 5.1  & 23 & 220 & 3.5  \\
    dragonbox & 16 & 200 & 4.7  & 30 & 400 & 4.8  & 19 & 170 & 3.2  \\
    ryu & 25 & 320 & 4.6  & 32 & 460 & 5.1  & 24 & 240 & 3.7  \\
    double\_conversion & 55 & 650 & 4.2  & 80 & 950 & 4.2  & 63 & 710 & 4.0  \\
    swiftDtoa & 22 & 290 & 4.7  & 34 & 490 & 5.1  & 29 & 300 & 3.6  \\
    std::to\_chars & 34 & 500 & 5.3  & 45 & 640 & 5.1  & 34 & 410 & 4.4  \\
    \bottomrule
  \end{tabular}\restartrowcolors
\end{table}

%% file: results/c7a.metal-48xl/AMD_EPYC_9R14_96-Core_Processor_g++_all_none.tex
\begin{table}
  \centering
  \caption{AMD EPYC 9R14 results (g++, 64-bit floats)}%
  \label{tab:amdepyc9r1496coreprocessorresults}
  \begin{tabular}{lccccccccc}
    \toprule
    \multirow{1}{*}{Name}   & \multicolumn{3}{c|}{mesh}  & \multicolumn{3}{c|}{canada} & \multicolumn{3}{c}{unit} \\
                            & {ns/f} & {ins/f} & {ins/c} & {ns/f} & {ins/f} & {ins/c}  & {ns/f} & {ins/f} & {ins/c} \\ \midrule
    dragon4 & 130 & 2300 & 4.9  & 260 & 4700 & 4.9  & 280 & 4900 & 4.8  \\
    fmt\_format & 46 & 610 & 3.6  & 83 & 1200 & 4.0  & 52 & 590 & 3.1  \\
    grisu3 & 21 & 310 & 3.9  & 49 & 680 & 3.8  & 42 & 520 & 3.4  \\
    grisu\_exact & 28 & 390 & 3.8  & 41 & 580 & 3.8  & 33 & 380 & 3.2  \\
    schubfach & 19 & 270 & 3.8  & 43 & 550 & 3.5  & 32 & 340 & 2.8  \\
    dragonbox & 17 & 270 & 4.4  & 32 & 450 & 3.8  & 24 & 250 & 2.8  \\
    ryu & 22 & 350 & 4.3  & 42 & 630 & 4.1  & 32 & 410 & 3.5  \\
    double\_conversion & 48 & 640 & 3.6  & 80 & 1100 & 3.6  & 70 & 840 & 3.2  \\
    swiftDtoa & 41 & 530 & 3.5  & 53 & 660 & 3.4  & 45 & 460 & 2.8  \\
    std::to\_chars & 32 & 510 & 4.3  & 53 & 860 & 4.4  & 46 & 640 & 3.8  \\
    \bottomrule
  \end{tabular}\restartrowcolors
\end{table}

%% file: results/c7a.metal-48xl/AMD_EPYC_9R14_96-Core_Processor_clang++_all_none.tex
\begin{table}
  \centering
  \caption{AMD EPYC 9R14 results (clang++, 64-bit floats)}%
  \label{tab:amdepyc9r1496coreprocessorresults}
  \begin{tabular}{lccccccccc}
    \toprule
    \multirow{1}{*}{Name}   & \multicolumn{3}{c|}{mesh}  & \multicolumn{3}{c|}{canada} & \multicolumn{3}{c}{unit} \\
                            & {ns/f} & {ins/f} & {ins/c} & {ns/f} & {ins/f} & {ins/c}  & {ns/f} & {ins/f} & {ins/c} \\ \midrule
    dragon4 & 120 & 2100 & 4.9  & 250 & 4400 & 4.7  & 250 & 4600 & 4.9  \\
    fmt\_format & 42 & 540 & 3.5  & 63 & 860 & 3.7  & 47 & 510 & 2.9  \\
    grisu3 & 21 & 290 & 3.8  & 50 & 630 & 3.4  & 44 & 470 & 2.9  \\
    grisu\_exact & 24 & 340 & 3.8  & 39 & 510 & 3.6  & 30 & 320 & 2.9  \\
    schubfach & 17 & 240 & 3.9  & 39 & 490 & 3.5  & 31 & 290 & 2.5  \\
    dragonbox & 18 & 250 & 3.8  & 31 & 420 & 3.6  & 24 & 230 & 2.6  \\
    ryu & 22 & 310 & 3.9  & 39 & 560 & 3.8  & 31 & 350 & 3.1  \\
    double\_conversion & 54 & 720 & 3.6  & 91 & 1200 & 3.6  & 78 & 970 & 3.4  \\
    swiftDtoa & 32 & 440 & 3.7  & 43 & 560 & 3.5  & 36 & 370 & 2.8  \\
    std::to\_chars & 32 & 510 & 4.4  & 52 & 850 & 4.4  & 45 & 640 & 3.9  \\
    \bottomrule
  \end{tabular}\restartrowcolors
\end{table}

%% file: results/c7a.metal-48xl/AMD_EPYC_9R14_96-Core_Processor_g++_all_s.tex
\begin{table}
  \centering
  \caption{AMD EPYC 9R14 results (g++, 32-bit floats)}%
  \label{tab:amdepyc9r1496coreprocessorresults}
  \begin{tabular}{lccccccccc}
    \toprule
    \multirow{1}{*}{Name}   & \multicolumn{3}{c|}{mesh}  & \multicolumn{3}{c|}{canada} & \multicolumn{3}{c}{unit} \\
                            & {ns/f} & {ins/f} & {ins/c} & {ns/f} & {ins/f} & {ins/c}  & {ns/f} & {ins/f} & {ins/c} \\ \midrule
    dragon4 & 99 & 1700 & 4.6  & 130 & 2400 & 5.0  & 150 & 2400 & 4.2  \\
    fmt\_format & 39 & 530 & 3.7  & 62 & 910 & 4.0  & 41 & 460 & 3.0  \\
    grisu\_exact & 25 & 330 & 3.6  & 38 & 490 & 3.5  & 30 & 290 & 2.6  \\
    schubfach & 17 & 240 & 3.8  & 32 & 460 & 3.9  & 24 & 250 & 2.8  \\
    dragonbox & 15 & 220 & 3.9  & 29 & 390 & 3.6  & 19 & 180 & 2.6  \\
    ryu & 28 & 400 & 3.9  & 35 & 520 & 4.0  & 28 & 320 & 3.1  \\
    double\_conversion & 46 & 640 & 3.8  & 71 & 920 & 3.5  & 57 & 690 & 3.3  \\
    swiftDtoa & 24 & 310 & 3.4  & 37 & 490 & 3.6  & 31 & 290 & 2.6  \\
    std::to\_chars & 37 & 560 & 4.1  & 47 & 680 & 4.0  & 38 & 460 & 3.3  \\
    \bottomrule
  \end{tabular}\restartrowcolors
\end{table}

%% file: results/c7a.metal-48xl/AMD_EPYC_9R14_96-Core_Processor_clang++_all_s.tex
\begin{table}
  \centering
  \caption{AMD EPYC 9R14 results (clang++, 32-bit floats)}%
  \label{tab:amdepyc9r1496coreprocessorresults}
  \begin{tabular}{lccccccccc}
    \toprule
    \multirow{1}{*}{Name}   & \multicolumn{3}{c|}{mesh}  & \multicolumn{3}{c|}{canada} & \multicolumn{3}{c}{unit} \\
                            & {ns/f} & {ins/f} & {ins/c} & {ns/f} & {ins/f} & {ins/c}  & {ns/f} & {ins/f} & {ins/c} \\ \midrule
    dragon4 & 87 & 1600 & 4.9  & 120 & 2300 & 5.0  & 130 & 2200 & 4.6  \\
    fmt\_format & 36 & 460 & 3.5  & 54 & 730 & 3.6  & 38 & 390 & 2.8  \\
    grisu\_exact & 23 & 300 & 3.6  & 35 & 460 & 3.6  & 29 & 270 & 2.5  \\
    schubfach & 16 & 220 & 3.6  & 31 & 430 & 3.8  & 23 & 220 & 2.6  \\
    dragonbox & 15 & 200 & 3.7  & 28 & 370 & 3.7  & 19 & 170 & 2.5  \\
    ryu & 27 & 380 & 3.8  & 34 & 480 & 3.8  & 28 & 290 & 2.8  \\
    double\_conversion & 55 & 730 & 3.6  & 80 & 1000 & 3.5  & 66 & 800 & 3.3  \\
    swiftDtoa & 23 & 310 & 3.7  & 34 & 490 & 3.9  & 31 & 310 & 2.7  \\
    std::to\_chars & 37 & 550 & 4.1  & 46 & 670 & 4.0  & 38 & 460 & 3.3  \\
    \bottomrule
  \end{tabular}\restartrowcolors
\end{table}

%% file: results/c6a.metal/AMD_EPYC_7R13_48-Core_Processor_g++_all_none.tex
\begin{table}
  \centering
  \caption{AMD EPYC 7R13 results (g++, 64-bit floats)}%
  \label{tab:amdepyc7r1348coreprocessorresults}
  \begin{tabular}{lccccccccc}
    \toprule
    \multirow{1}{*}{Name}   & \multicolumn{3}{c|}{mesh}  & \multicolumn{3}{c|}{canada} & \multicolumn{3}{c}{unit} \\
                            & {ns/f} & {ins/f} & {ins/c} & {ns/f} & {ins/f} & {ins/c}  & {ns/f} & {ins/f} & {ins/c} \\ \midrule
    dragon4 & 130 & 2300 & 4.6  & 280 & 4600 & 4.5  & 290 & 4900 & 4.5  \\
    fmt\_format & 48 & 610 & 3.4  & 88 & 1100 & 3.5  & 54 & 590 & 3.0  \\
    grisu3 & 23 & 310 & 3.6  & 52 & 680 & 3.5  & 44 & 520 & 3.2  \\
    grisu\_exact & 29 & 390 & 3.6  & 44 & 570 & 3.5  & 33 & 380 & 3.1  \\
    schubfach & 20 & 270 & 3.5  & 46 & 550 & 3.2  & 34 & 340 & 2.7  \\
    dragonbox & 18 & 270 & 3.9  & 34 & 450 & 3.6  & 24 & 250 & 2.8  \\
    ryu & 24 & 350 & 3.8  & 43 & 630 & 3.9  & 32 & 410 & 3.5  \\
    double\_conversion & 54 & 630 & 3.2  & 94 & 1100 & 3.0  & 80 & 830 & 2.8  \\
    swiftDtoa & 43 & 530 & 3.3  & 55 & 660 & 3.2  & 46 & 460 & 2.7  \\
    std::to\_chars & 37 & 520 & 3.8  & 57 & 860 & 4.1  & 48 & 640 & 3.6  \\
    \bottomrule
  \end{tabular}\restartrowcolors
\end{table}

%% file: results/c6a.metal/AMD_EPYC_7R13_48-Core_Processor_clang++_all_none.tex
\begin{table}
  \centering
  \caption{AMD EPYC 7R13 results (clang++, 64-bit floats)}%
  \label{tab:amdepyc7r1348coreprocessorresults}
  \begin{tabular}{lccccccccc}
    \toprule
    \multirow{1}{*}{Name}   & \multicolumn{3}{c|}{mesh}  & \multicolumn{3}{c|}{canada} & \multicolumn{3}{c}{unit} \\
                            & {ns/f} & {ins/f} & {ins/c} & {ns/f} & {ins/f} & {ins/c}  & {ns/f} & {ins/f} & {ins/c} \\ \midrule
    dragon4 & 120 & 2100 & 4.6  & 270 & 4400 & 4.4  & 270 & 4600 & 4.5  \\
    fmt\_format & 45 & 540 & 3.2  & 67 & 840 & 3.4  & 48 & 510 & 2.8  \\
    grisu3 & 22 & 290 & 3.5  & 53 & 630 & 3.2  & 44 & 470 & 2.9  \\
    grisu\_exact & 26 & 340 & 3.5  & 41 & 510 & 3.4  & 30 & 320 & 2.9  \\
    schubfach & 18 & 240 & 3.5  & 43 & 490 & 3.1  & 30 & 290 & 2.6  \\
    dragonbox & 20 & 250 & 3.3  & 33 & 420 & 3.5  & 24 & 230 & 2.6  \\
    ryu & 23 & 310 & 3.6  & 40 & 550 & 3.8  & 30 & 350 & 3.1  \\
    double\_conversion & 58 & 720 & 3.3  & 100 & 1200 & 3.1  & 86 & 960 & 3.0  \\
    swiftDtoa & 34 & 440 & 3.4  & 44 & 550 & 3.4  & 37 & 380 & 2.7  \\
    std::to\_chars & 36 & 510 & 3.8  & 57 & 850 & 4.0  & 47 & 640 & 3.7  \\
    \bottomrule
  \end{tabular}\restartrowcolors
\end{table}

%% file: results/c6a.metal/AMD_EPYC_7R13_48-Core_Processor_g++_all_s.tex
\begin{table}
  \centering
  \caption{AMD EPYC 7R13 results (g++, 32-bit floats)}%
  \label{tab:amdepyc7r1348coreprocessorresults}
  \begin{tabular}{lccccccccc}
    \toprule
    \multirow{1}{*}{Name}   & \multicolumn{3}{c|}{mesh}  & \multicolumn{3}{c|}{canada} & \multicolumn{3}{c}{unit} \\
                            & {ns/f} & {ins/f} & {ins/c} & {ns/f} & {ins/f} & {ins/c}  & {ns/f} & {ins/f} & {ins/c} \\ \midrule
    dragon4 & 100 & 1700 & 4.4  & 150 & 2400 & 4.4  & 160 & 2400 & 4.1  \\
    fmt\_format & 44 & 530 & 3.3  & 69 & 890 & 3.5  & 43 & 460 & 2.9  \\
    grisu\_exact & 27 & 330 & 3.2  & 40 & 480 & 3.3  & 30 & 290 & 2.6  \\
    schubfach & 19 & 230 & 3.3  & 37 & 460 & 3.3  & 24 & 250 & 2.8  \\
    dragonbox & 17 & 210 & 3.4  & 31 & 380 & 3.3  & 19 & 180 & 2.5  \\
    ryu & 29 & 400 & 3.7  & 38 & 520 & 3.6  & 28 & 320 & 3.0  \\
    double\_conversion & 57 & 640 & 3.0  & 80 & 910 & 3.1  & 64 & 690 & 2.9  \\
    swiftDtoa & 27 & 310 & 3.1  & 40 & 490 & 3.3  & 31 & 290 & 2.5  \\
    std::to\_chars & 41 & 560 & 3.7  & 51 & 680 & 3.6  & 41 & 460 & 3.0  \\
    \bottomrule
  \end{tabular}\restartrowcolors
\end{table}

%% file: results/c6a.metal/AMD_EPYC_7R13_48-Core_Processor_clang++_all_s.tex
\begin{table}
  \centering
  \caption{AMD EPYC 7R13 results (clang++, 32-bit floats)}%
  \label{tab:amdepyc7r1348coreprocessorresults}
  \begin{tabular}{lccccccccc}
    \toprule
    \multirow{1}{*}{Name}   & \multicolumn{3}{c|}{mesh}  & \multicolumn{3}{c|}{canada} & \multicolumn{3}{c}{unit} \\
                            & {ns/f} & {ins/f} & {ins/c} & {ns/f} & {ins/f} & {ins/c}  & {ns/f} & {ins/f} & {ins/c} \\ \midrule
    dragon4 & 94 & 1600 & 4.5  & 140 & 2300 & 4.4  & 140 & 2200 & 4.2  \\
    fmt\_format & 39 & 460 & 3.2  & 58 & 730 & 3.4  & 40 & 390 & 2.7  \\
    grisu\_exact & 23 & 300 & 3.4  & 36 & 460 & 3.4  & 29 & 270 & 2.5  \\
    schubfach & 18 & 220 & 3.3  & 37 & 430 & 3.2  & 23 & 220 & 2.6  \\
    dragonbox & 16 & 200 & 3.3  & 31 & 370 & 3.2  & 19 & 170 & 2.5  \\
    ryu & 28 & 380 & 3.6  & 36 & 480 & 3.6  & 27 & 290 & 2.8  \\
    double\_conversion & 60 & 730 & 3.2  & 87 & 1000 & 3.2  & 72 & 800 & 3.0  \\
    swiftDtoa & 25 & 310 & 3.4  & 37 & 490 & 3.5  & 32 & 310 & 2.6  \\
    std::to\_chars & 40 & 550 & 3.7  & 50 & 670 & 3.6  & 41 & 460 & 3.0  \\
    \bottomrule
  \end{tabular}\restartrowcolors
\end{table}

%% file: results/c5a.24xlarge/AMD_EPYC_7R32_g++_all_none.tex
\begin{table}
  \centering
  \caption{AMD EPYC 7R32 results (g++, 64-bit floats)}%
  \label{tab:amdepyc7r32results}
  \begin{tabular}{lccccccccc}
    \toprule
    \multirow{1}{*}{Name}   & \multicolumn{3}{c|}{mesh}  & \multicolumn{3}{c|}{canada} & \multicolumn{3}{c}{unit} \\
                            & {ns/f} & {ins/f} & {ins/c} & {ns/f} & {ins/f} & {ins/c}  & {ns/f} & {ins/f} & {ins/c} \\ \midrule
    dragon4 & 170 & 2300 & 4.0  & 370 & 4600 & 3.9  & 370 & 4900 & 4.0  \\
    fmt\_format & 63 & 610 & 3.0  & 130 & 1100 & 2.7  & 67 & 590 & 2.7  \\
    grisu3 & 31 & 310 & 3.0  & 72 & 680 & 2.9  & 61 & 520 & 2.6  \\
    grisu\_exact & 38 & 390 & 3.1  & 58 & 570 & 3.1  & 41 & 380 & 2.8  \\
    schubfach & 27 & 270 & 3.0  & 64 & 550 & 2.7  & 41 & 340 & 2.6  \\
    dragonbox & 24 & 270 & 3.4  & 49 & 450 & 2.8  & 30 & 250 & 2.6  \\
    ryu & 33 & 350 & 3.2  & 58 & 630 & 3.3  & 40 & 410 & 3.2  \\
    double\_conversion & 69 & 630 & 2.8  & 120 & 1100 & 2.6  & 99 & 830 & 2.6  \\
    swiftDtoa & 56 & 530 & 2.9  & 74 & 660 & 2.7  & 59 & 460 & 2.4  \\
    std::to\_chars & 54 & 520 & 3.0  & 80 & 860 & 3.3  & 60 & 640 & 3.3  \\
    \bottomrule
  \end{tabular}\restartrowcolors
\end{table}

%% file: results/c5a.24xlarge/AMD_EPYC_7R32_clang++_all_none.tex
\begin{table}
  \centering
  \caption{AMD EPYC 7R32 results (clang++, 64-bit floats)}%
  \label{tab:amdepyc7r32results}
  \begin{tabular}{lccccccccc}
    \toprule
    \multirow{1}{*}{Name}   & \multicolumn{3}{c|}{mesh}  & \multicolumn{3}{c|}{canada} & \multicolumn{3}{c}{unit} \\
                            & {ns/f} & {ins/f} & {ins/c} & {ns/f} & {ins/f} & {ins/c}  & {ns/f} & {ins/f} & {ins/c} \\ \midrule
    dragon4 & 170 & 1900 & 3.5  & 340 & 4000 & 3.6  & 370 & 4200 & 3.5  \\
    fmt\_format & 61 & 540 & 2.7  & 94 & 830 & 2.8  & 61 & 500 & 2.6  \\
    grisu3 & 28 & 290 & 3.1  & 70 & 620 & 2.8  & 61 & 470 & 2.4  \\
    grisu\_exact & 34 & 340 & 3.1  & 57 & 510 & 2.8  & 38 & 320 & 2.6  \\
    schubfach & 23 & 240 & 3.1  & 53 & 490 & 2.9  & 36 & 290 & 2.4  \\
    dragonbox & 27 & 250 & 2.9  & 51 & 420 & 2.6  & 29 & 230 & 2.4  \\
    ryu & 31 & 310 & 3.1  & 51 & 550 & 3.4  & 39 & 350 & 2.7  \\
    double\_conversion & 78 & 710 & 2.8  & 130 & 1200 & 2.7  & 110 & 970 & 2.7  \\
    swiftDtoa & 42 & 440 & 3.2  & 56 & 550 & 3.0  & 45 & 370 & 2.5  \\
    std::to\_chars & 52 & 510 & 3.0  & 77 & 850 & 3.4  & 60 & 640 & 3.3  \\
    \bottomrule
  \end{tabular}\restartrowcolors
\end{table}

%% file: results/c5a.24xlarge/AMD_EPYC_7R32_g++_all_s.tex
\begin{table}
  \centering
  \caption{AMD EPYC 7R32 results (g++, 32-bit floats)}%
  \label{tab:amdepyc7r32results}
  \begin{tabular}{lccccccccc}
    \toprule
    \multirow{1}{*}{Name}   & \multicolumn{3}{c|}{mesh}  & \multicolumn{3}{c|}{canada} & \multicolumn{3}{c}{unit} \\
                            & {ns/f} & {ins/f} & {ins/c} & {ns/f} & {ins/f} & {ins/c}  & {ns/f} & {ins/f} & {ins/c} \\ \midrule
    dragon4 & 140 & 1700 & 3.8  & 200 & 2400 & 3.8  & 190 & 2400 & 3.8  \\
    fmt\_format & 56 & 530 & 2.9  & 110 & 890 & 2.6  & 56 & 460 & 2.5  \\
    grisu\_exact & 35 & 330 & 2.8  & 56 & 480 & 2.7  & 40 & 290 & 2.2  \\
    schubfach & 24 & 230 & 3.0  & 50 & 460 & 2.8  & 31 & 250 & 2.5  \\
    dragonbox & 22 & 210 & 3.0  & 47 & 380 & 2.5  & 24 & 180 & 2.3  \\
    ryu & 35 & 400 & 3.5  & 54 & 520 & 2.9  & 34 & 320 & 2.8  \\
    double\_conversion & 70 & 640 & 2.8  & 100 & 910 & 2.7  & 77 & 690 & 2.8  \\
    swiftDtoa & 34 & 310 & 2.8  & 54 & 490 & 2.8  & 39 & 290 & 2.3  \\
    std::to\_chars & 58 & 560 & 3.0  & 76 & 680 & 2.8  & 57 & 460 & 2.5  \\
    \bottomrule
  \end{tabular}\restartrowcolors
\end{table}

%% file: results/c5a.24xlarge/AMD_EPYC_7R32_clang++_all_s.tex
\begin{table}
  \centering
  \caption{AMD EPYC 7R32 results (clang++, 32-bit floats)}%
  \label{tab:amdepyc7r32results}
  \begin{tabular}{lccccccccc}
    \toprule
    \multirow{1}{*}{Name}   & \multicolumn{3}{c|}{mesh}  & \multicolumn{3}{c|}{canada} & \multicolumn{3}{c}{unit} \\
                            & {ns/f} & {ins/f} & {ins/c} & {ns/f} & {ins/f} & {ins/c}  & {ns/f} & {ins/f} & {ins/c} \\ \midrule
    dragon4 & 130 & 1500 & 3.4  & 190 & 2100 & 3.4  & 190 & 2000 & 3.2  \\
    fmt\_format & 51 & 460 & 2.8  & 77 & 740 & 3.0  & 48 & 390 & 2.5  \\
    grisu\_exact & 30 & 300 & 3.0  & 48 & 460 & 2.9  & 36 & 270 & 2.3  \\
    schubfach & 23 & 220 & 2.8  & 43 & 430 & 3.0  & 33 & 220 & 2.1  \\
    dragonbox & 21 & 200 & 3.0  & 44 & 370 & 2.6  & 24 & 170 & 2.2  \\
    ryu & 35 & 380 & 3.3  & 45 & 480 & 3.3  & 34 & 290 & 2.6  \\
    double\_conversion & 78 & 720 & 2.8  & 120 & 1000 & 2.7  & 93 & 790 & 2.6  \\
    swiftDtoa & 32 & 310 & 3.0  & 47 & 490 & 3.2  & 40 & 310 & 2.4  \\
    std::to\_chars & 56 & 550 & 3.0  & 69 & 670 & 3.0  & 58 & 460 & 2.4  \\
    \bottomrule
  \end{tabular}\restartrowcolors
\end{table}

%% file: results/apple_m4/Apple_M4_Max_clang++_all_none.tex
\begin{table}
  \centering
  \caption{Apple M4 Max results (clang++, 64-bit floats)}%
  \label{tab:applem4maxresults}
  \begin{tabular}{lccccccccc}
    \toprule
    \multirow{1}{*}{Name}   & \multicolumn{3}{c|}{mesh}  & \multicolumn{3}{c|}{canada} & \multicolumn{3}{c}{unit} \\
                            & {ns/f} & {ins/f} & {ins/c} & {ns/f} & {ins/f} & {ins/c}  & {ns/f} & {ins/f} & {ins/c} \\ \midrule
    dragon4 & 69 & 1500 & 5.3  & 150 & 3000 & 4.8  & 170 & 3300 & 4.6  \\
    fmt\_format & 22 & 530 & 5.4  & 29 & 640 & 5.0  & 30 & 510 & 3.8  \\
    grisu3 & 10 & 260 & 5.6  & 24 & 440 & 4.2  & 26 & 470 & 4.0  \\
    grisu\_exact & 11 & 320 & 6.3  & 15 & 340 & 5.1  & 18 & 340 & 4.2  \\
    schubfach & 7.2 & 210 & 6.4  & 12 & 310 & 5.9  & 14 & 290 & 4.7  \\
    dragonbox & 7.7 & 220 & 6.6  & 9.5 & 240 & 5.6  & 12 & 230 & 4.2  \\
    ryu & 9.9 & 270 & 6.0  & 12 & 330 & 6.3  & 13 & 310 & 5.4  \\
    double\_conversion & 26 & 640 & 5.5  & 42 & 910 & 5.1  & 43 & 880 & 4.8  \\
    swiftDtoa & 14 & 390 & 6.0  & 16 & 360 & 5.1  & 20 & 390 & 4.4  \\
    std::to\_chars & 13 & 350 & 5.8  & 15 & 440 & 6.6  & 16 & 410 & 5.6  \\
    \bottomrule
  \end{tabular}\restartrowcolors
\end{table}

%% file: results/apple_m4/Apple_M4_Max_clang++_all_s.tex
\begin{table}
  \centering
  \caption{Apple M4 Max results (clang++, 32-bit floats)}%
  \label{tab:applem4maxresults}
  \begin{tabular}{lccccccccc}
    \toprule
    \multirow{1}{*}{Name}   & \multicolumn{3}{c|}{mesh}  & \multicolumn{3}{c|}{canada} & \multicolumn{3}{c}{unit} \\
                            & {ns/f} & {ins/f} & {ins/c} & {ns/f} & {ins/f} & {ins/c}  & {ns/f} & {ins/f} & {ins/c} \\ \midrule
    dragon4 & 53 & 1200 & 5.3  & 80 & 1600 & 4.6  & 88 & 1600 & 4.4  \\
    fmt\_format & 18 & 450 & 5.7  & 22 & 540 & 5.7  & 20 & 410 & 4.5  \\
    grisu\_exact & 8.4 & 240 & 6.4  & 12 & 250 & 4.6  & 15 & 250 & 3.6  \\
    schubfach & 6.7 & 190 & 6.1  & 9.2 & 240 & 5.9  & 12 & 230 & 4.3  \\
    dragonbox & 6.3 & 170 & 6.1  & 8.9 & 180 & 4.6  & 9.9 & 170 & 3.8  \\
    ryu & 10 & 300 & 6.5  & 10 & 250 & 5.3  & 13 & 240 & 4.1  \\
    double\_conversion & 28 & 640 & 5.2  & 34 & 760 & 5.2  & 34 & 730 & 5.0  \\
    swiftDtoa & 11 & 290 & 5.7  & 12 & 300 & 5.6  & 19 & 320 & 3.8  \\
    std::to\_chars & 15 & 420 & 6.4  & 15 & 400 & 6.0  & 19 & 390 & 4.6  \\
    \bottomrule
  \end{tabular}\restartrowcolors
\end{table}